\documentclass[aps,prd,superscriptaddress,showpacs,preprint,amsmath,amssymb]{revtex4}
\usepackage{graphicx, bm}
\usepackage[usenames]{color}

\begin{document}

\draft
\title{Anomalous quartic $WW\gamma\gamma$ couplings in $ep$ collisions at the LHeC and the FCC-he}

\author{V. Ari\footnote{vari@science.ankara.edu.tr}}
\affiliation{\small Department of Physics, Ankara University, Turkey.\\}

\author{E. Gurkanli\footnote{egurkanli@sinop.edu.tr}}
\affiliation{\small Department of Physics, Sinop University, Turkey.\\}

\author{ A. Guti\'errez-Rodr\'{\i}guez\footnote{alexgu@fisica.uaz.edu.mx}}
\affiliation{\small Facultad de F\'{\i}sica, Universidad Aut\'onoma de Zacatecas\\
         Apartado Postal C-580, 98060 Zacatecas, M\'exico.\\}

\author{ M. A. Hern\'andez-Ru\'{\i}z\footnote{mahernan@uaz.edu.mx}}
\affiliation{\small Unidad Acad\'emica de Ciencias Qu\'{\i}micas, Universidad Aut\'onoma de Zacatecas\\
         Apartado Postal C-585, 98060 Zacatecas, M\'exico.\\}

\author{M. K\"{o}ksal\footnote{mkoksal@cumhuriyet.edu.tr}}
\affiliation{\small Deparment of Optical Engineering, Sivas Cumhuriyet University, 58140, Sivas, Turkey.\\}

\date{\today}

\begin{abstract}

We conducted a study on measuring $W^+W^-$ production and on the sensitivity limits at $95\%$ Confidence Level on thirteen anomalous
couplings obtained by dimension-8 operators which are related to the anomalous quartic $WW\gamma\gamma$ couplings. We consider
the main $e^-p \to e^-\gamma^*\gamma^*p \to e^-W^+W^-p$ reaction with the sub-process $\gamma^*\gamma^* \to W^+W^-$
at the Large Hadron electron Collider (LHeC) and the Future Circular Collider-hadron electron (FCC-he). For the LHeC, energies of the
$e^-$ beams are taken to be $E_e =60$ and 140 GeV and the energy of the $p$ beams is taken to be $E_p = 7$ TeV. For the FCC-he, energies
of the $e^-$ beams are taken to be $E_e =60$ and 140 GeV and the energy of the $p$ beams is taken to be $E_p = 50$ TeV, respectively.
It is interesting to notice that the LHeC and the FCC-he will lead to model-independent limits on the anomalous quartic $WW\gamma\gamma$
couplings which are one order of magnitude stringent than the CMS Collaboration limits, in addition to being competitive with other limits
reported in the literature.

\end{abstract}


\vspace{5mm}

\maketitle

\section{Introduction}

In the Standard Model (SM) of Elementary Particles Physics \cite{SM1,SM2,SM3}, electroweak gauge bosons ($W^\pm, Z, \gamma$)
are introduced to preserve the local gauge symmetry $SU(2)_L\times U(1)_Y$. As a consequence, there is a universality among
the couplings of fermions to the gauge boson, the three gauge bosons, and the four gauge bosons. This universality forms the
basis of the success of the SM \cite{Hagiwara0}. It is worth mentioning that, so far the fermion-gauge-boson couplings were
tested precisely at various colliders, however, the direct measurement of the self couplings of the gauge bosons is not precise
enough. For these reasons, it is very important to search and propose model-independent study to be able to measure with great
precision the anomalous Quartic-Gauge-Boson Couplings (aQGC) of the $W^\pm$ bosons. In addition, on this subject, the aQGC
$WW\gamma\gamma$, $WWZ\gamma$, $WWZZ$, $WWWW$ provide a window into one of the most important problems
in particle physics; the understanding of electroweak symmetry breaking.

The aQGC have been studied by LEP \cite{Barate,Abreu,Acciarri,Abbiendi} and Fermilab Tevatron \cite{Gounder,Abbott}. Presently,
the aQGC are being probed extensively by ATLAS and CMS Collaborations at the Large Hadron Collider (LHC) \cite{Chatrchyan,Aaboud}.
In addition, different theoretical and phenomenological groups have been carried several studies on the aQGC in a different context
\cite{Belanger,Stirling,Leil,Bervan,Chong,Koksal,Chen,Stirling1,Aaboud,Atag,Eboli,Eboli1,Sahin,Koksal1,Chapon,Koksal2,Senol,Koksal3,Yang,Eboli2,
Eboli4,Bell,Ahmadov,Schonherr,Wen,Ye,Eboli3,Perez,Sahin1,Senol1,Baldenegro,Fichet,Fichet1,Pierzchala,Gutierrez}. However, the possibility of high-energy photon interactions in present and future colliders such as the LHC, the Large Hadron electron Collider (LHeC) and the Future Circular Collider-hadron electron (FCC-he) \cite{FCChe,Fernandez,Fernandez1,Fernandez2,Huan,Acar} opens up the possibility of new research on the aQGC. These present and future projects offer a unique possibility for a novel and complementary research of the aQGC through the two-photon associated production and a pair of $W^\pm$ bosons via the
process $e^-p \to e^-\gamma^*\gamma^*p \to e^-W^+W^-p$ with the corresponding sub-process $\gamma^*\gamma^* \to W^+W^-$.

This paper is organized as follows. In Sect. II, a brief review of the operators in our effective Lagrangian is provided.
In Sect. III, we derive limits for the aQGC at the LHeC and the FCC-he. In Sect. IV, we present our conclusions.

\section{Theoretical approach of the aQGC $WW\gamma\gamma$}

Exploring the process $e^-p \to e^-\gamma^*\gamma^*p \to e^-W^+W^-p$, as well as the aQGC $WW\gamma\gamma$ through precise
measurements at the present and future facilities are quite challenging. Thus, once such the aQGC $WW\gamma\gamma$ are
measured with great precision, it must be a strong indication of new physics Beyond the Standard Model (BSM). Considering this
situation, we performed a model-independent study of possible anomalous quartic $WW\gamma\gamma$ couplings, in the
effective Lagrangian framework.

If baryon and lepton numbers are conserved, only operators with even dimension can appear in the effective field theory. Hence, firstly, the largest new physics contribution is anticipated from dimension-6 operators. Three CP conserving dimension-6 operators are

\begin{eqnarray}
O_{WWW}=Tr[W_{\mu\nu}W^{\nu\rho}W^{\mu}_{\rho}]\\
O_{W}=(D_{\mu}\Phi)^{\dagger}W^{\mu\nu}(D_{\nu}\Phi)\\
O_{B}=(D_{\mu}\Phi)^{\dagger}B^{\mu\nu}(D_{\nu}\Phi)\\
\end{eqnarray}

and two CP violating dimension-6 operators are

\begin{eqnarray}
O_{\tilde{W}WW}=Tr[\tilde{W}_{\mu\nu}W^{\nu\rho}W^{\mu}_{\rho}]\\
O_{\tilde{W}}=(D_{\mu}\Phi)^{\dagger}\tilde{W}^{\mu\nu}(D_{\nu}\Phi)\\
\end{eqnarray}
where $\Phi$ is the Higgs doublet field.

$O_{WWW}$, $O_{W}$ and $O_{\tilde{W}WW}$ operators affect the triple gauge couplings ($WW\gamma$, $WWZ$) and the quartic gauge couplings ($ WWWW$, $WW \gamma \gamma$, $WW \gamma Z$ and $WWZZ$). Hence, we find out that the dimension-6 operators giving rise to the quartic gauge couplings also exhibit the triple gauge couplings.

In order to separate the effects of the quartic gauge couplings we shall consider effective operators that lead to the quartic gauge couplings without a triple gauge couplings associated to them. Also, not all possible QGCs are generated by dimension-6 operators. The lowest dimension operator that leads to quartic interactions but does not exhibit two or three weak gauge boson vertices is of dimension-8. For this reason, genuine quartic vertices are of dimension-8 or higher. The idea behind using dimension-8 operators for the quartic gauge couplings is that the anomalous quartic gauge couplings to study these couplings without having any theoretical prejudice about their size. Especially, vector boson scattering processes are widely recognized as the best laboratory to study dimension-8 operators, which modify only the $V V V V$ quartic couplings.

The corresponding interaction effective Lagrangian comes from several $SU(2)\times U(1)$ invariant dimension-8 effective operators
that modify the interactions among electroweak gauge bosons is given by \cite{Degrande}

\begin{equation}
{\cal L}_{eff}=  \sum_{j=1}^2\frac{f_{S, j}}{\Lambda^4}O_{S, j} + \sum_{j=0}^{9}\frac{f_{T, j}}{\Lambda^4}O_{T, j}
+ \sum_{j=0}^{7}\frac{f_{M, j}}{\Lambda^4}O_{M, j}.
\end{equation}

In Eq. (8), there are 18 different operators that define the aQGC, and the indices $S$, $T$ and $M$ of the couplings represent
three classes of genuine aQGC operators. For the first class of these operators, there are two independent operators
below

\begin{eqnarray}
O_{S, 0}&=&[(D_\mu\Phi)^\dagger (D_\nu\Phi)]\times [(D^\mu\Phi)^\dagger (D^\nu\Phi)],  \\
O_{S, 1}&=&[(D_\mu\Phi)^\dagger (D^\mu\Phi)]\times [(D_\nu\Phi)^\dagger (D^\nu\Phi)],
\end{eqnarray}

\noindent where $D_\mu$ is the covariant derivative and $\Phi$ denotes the Higgs double field. $O_{S, 0}$ and
$O_{S, 1}$ operators in Eqs. (9)-(10) contain the quartic $WWWW$, $WWZZ$ and $ZZZZ$ couplings. These operators are also known as scalar
operators.

Another alternative way to generate the aQGC is through operators containing $D^\mu\Phi$ as well as two field strength tensors,
that is

\begin{eqnarray}
O_{M, 0}&=&Tr[W_{\mu\nu} W^{\mu\nu}]\times [(D_\beta\Phi)^\dagger (D^\beta\Phi)],  \\
O_{M, 1}&=&Tr[W_{\mu\nu} W^{\nu\beta}]\times [(D_\beta\Phi)^\dagger (D^\mu\Phi)],  \\
O_{M, 2}&=&[B_{\mu\nu} B^{\mu\nu}]\times [(D_\beta\Phi)^\dagger (D^\beta\Phi)],  \\
O_{M, 3}&=&[B_{\mu\nu} B^{\nu\beta}]\times [(D_\beta\Phi)^\dagger (D^\mu\Phi)],  \\
O_{M, 4}&=&[(D_\mu\Phi)^\dagger W_{\beta\nu} (D^\mu\Phi)]\times B^{\beta\nu},  \\
O_{M, 5}&=&[(D_\mu\Phi)^\dagger W_{\beta\nu} (D^\nu\Phi)]\times B^{\beta\mu},  \\
O_{M, 6}&=&[(D_\mu\Phi)^\dagger W_{\beta\nu} W^{\beta\nu} (D^\mu\Phi)],  \\
O_{M, 7}&=&[(D_\mu\Phi)^\dagger W_{\beta\nu} W^{\beta\mu} (D^\nu\Phi)]
\end{eqnarray}

\noindent which are known as mixed operators.

In the case, when the aQGC contain only four field strength tensors, the structure of the operators is represented by

\begin{eqnarray}
O_{T, 0}&=&Tr[W_{\mu\nu} W^{\mu\nu}]\times Tr[W_{\alpha\beta}W^{\alpha\beta}],  \\
O_{T, 1}&=&Tr[W_{\alpha\nu} W^{\mu\beta}]\times Tr[W_{\mu\beta}W^{\alpha\nu}],  \\
O_{T, 2}&=&Tr[W_{\alpha\mu} W^{\mu\beta}]\times Tr[W_{\beta\nu}W^{\nu\alpha}],  \\
O_{T, 5}&=&Tr[W_{\mu\nu} W^{\mu\nu}]\times B_{\alpha\beta}B^{\alpha\beta},  \\
O_{T, 6}&=&Tr[W_{\alpha\nu} W^{\mu\beta}]\times B_{\mu\beta}B^{\alpha\nu},  \\
O_{T, 7}&=&Tr[W_{\alpha\mu} W^{\mu\beta}]\times B_{\beta\nu}B^{\nu\alpha},  \\
O_{T, 8}&=&B_{\mu\nu} B^{\mu\nu}B_{\alpha\beta}B^{\alpha\beta},  \\
O_{T, 9}&=&B_{\alpha\mu} B^{\mu\beta}B_{\beta\nu}B^{\nu\alpha}.
\end{eqnarray}

\noindent They are called transverse operators.

Also, the LEP constraints on the $WW\gamma \gamma$ vertices defined in terms of the anomalous $a_{0}/\Lambda^{2}$ and $a_{c}/\Lambda^{2}$ couplings
can be translated into limits on $f_{M,0}-f_{M,7}$. The genuine anomalous quartic couplings involving two photons have been introduced as follows
\cite{Baak}

\begin{eqnarray}
\frac{f_{M, 0}}{\Lambda^2}&=&\frac{a_0}{\Lambda^2}\frac{1}{g^2v^2},  \\
\frac{f_{M, 1}}{\Lambda^2}&=&-\frac{a_c}{\Lambda^2}\frac{1}{g^2v^2},  \\
\frac{f_{M, 0}}{\Lambda^2}&=&\frac{f_{M, 2}}{2}= \frac{f_{M, 6}}{2},    \\
\frac{f_{M, 1}}{\Lambda^2}&=&\frac{f_{M, 3}}{2}= -\frac{f_{M, 5}}{2}=\frac{f_{M, 7}}{2}.
\end{eqnarray}

Next, we present Table I which shows the experimental limits on the aQGC $f_{M, i}$ and $f_{T, i}$ that are set at $95\%$ Confidence Level (C.L.)
by the CMS Collaboration at the LHC via $pp \to p\gamma^*\gamma^*p \to pWWp$ \cite{CMS} through the sub-process $\gamma^*\gamma^* \to W^+W^-$ and $pp \to W\gamma jj$ \cite{CMS1} at $\sqrt{s}=8$ TeV and ${\cal L}=19.7$ $\rm fb^{-1}$, respectively. The limits on $f_{M, i}$ and $f_{T, i}$ given in Table I are of interest for the study that we carry out in this paper.

\begin{table} [ht]
\caption{Summary of experimental limits on aQGC at the $95\%$ C. L.  by the CMS Collaboration at the LHC via
$pp \to p\gamma^*\gamma^* \to pWWp$ \cite{CMS} and $pp \to W\gamma jj$ \cite{CMS1} at $\sqrt{s}=8$ TeV
and ${\cal L}=19.7$ $\rm fb^{-1}$.}
\begin{tabular}{|c|c|c|c|}
\hline\hline
{\bf Dimension-8 aQGC parameter} & {\bf Process $pp \to p\gamma^*\gamma^*p \to pWWp$}     & {\bf C. L.}    & {\bf Reference}    \\
\hline \hline
$f_{M,0}/\Lambda^{4}$ $(\rm TeV^{-4})$    & $-4.2 < f_{M,0}/\Lambda^{4} < 4.2$     & $95\%$   & \cite{CMS}    \\
\hline
$f_{M,1}/\Lambda^{4}$ $(\rm TeV^{-4})$    & $-16 < f_{M,1}/\Lambda^{4} < 16$       & $95\%$   & \cite{CMS}     \\
\hline
$f_{M,2}/\Lambda^{4}$ $(\rm TeV^{-4})$    & $-2.1 < f_{M,2}/\Lambda^{4} < 2.1$     & $95\%$   & \cite{CMS}     \\
\hline
$f_{M,3}/\Lambda^{4}$ $(\rm TeV^{-4})$    & $-7.8 < f_{M,3}/\Lambda^{4} < 7.8$     & $95\%$   & \cite{CMS}     \\
\hline\hline
{\bf Dimension-8 aQGC parameter} & {\bf Process $pp \to W\gamma jj$}               & {\bf C. L.}    & {\bf Reference}    \\
\hline \hline
$f_{M,4}/\Lambda^{4}$ $(\rm TeV^{-4})$    & $-40 < f_{M,4}/\Lambda^{4} < 40$       & $95\%$   & \cite{CMS1}     \\
\hline
$f_{M,5}/\Lambda^{4}$ $(\rm TeV^{-4})$    & $-65 < f_{M,5}/\Lambda^{4} < 65$       & $95\%$   & \cite{CMS1}     \\
\hline
$f_{M,6}/\Lambda^{4}$ $(\rm TeV^{-4})$    & $-129 < f_{M,6}/\Lambda^{4} < 129$     & $95\%$   & \cite{CMS1}     \\
\hline
$f_{M,7}/\Lambda^{4}$ $(\rm TeV^{-4})$    & $-164 < f_{M,7}/\Lambda^{4} < 162$     & $95\%$   & \cite{CMS1}     \\
\hline
$f_{T,0}/\Lambda^{4}$ $(\rm TeV^{-4})$    & $-5.4 < f_{T,0}/\Lambda^{4} < 5.6$     & $95\%$   & \cite{CMS1}     \\
\hline
$f_{T,1}/\Lambda^{4}$ $(\rm TeV^{-4})$    & $-3.7 < f_{T,1}/\Lambda^{4} < 4.0$     & $95\%$   & \cite{CMS1}     \\
\hline
$f_{T,2}/\Lambda^{4}$ $(\rm TeV^{-4})$    & $-11 < f_{T,2}/\Lambda^{4} < 12$       & $95\%$   & \cite{CMS1}     \\
\hline
$f_{T,5}/\Lambda^{4}$ $(\rm TeV^{-4})$    & $-3.8 < f_{T,5}/\Lambda^{4} < 3.8$     & $95\%$   & \cite{CMS1}     \\
\hline
$f_{T,6}/\Lambda^{4}$ $(\rm TeV^{-4})$    & $-2.8 < f_{T,6}/\Lambda^{4} < 3.0$     & $95\%$   & \cite{CMS1}     \\
\hline
$f_{T,7}/\Lambda^{4}$ $(\rm TeV^{-4})$    & $-7.3 < f_{T,0}/\Lambda^{4} < 7.7$     & $95\%$   & \cite{CMS1}     \\
\hline
\end{tabular}
\end{table}

\section{Cross-section of the process $e^-p \to e^-\gamma^*\gamma^*p \to e^-W^+W^-p$ and limits on the aQGC
at the LHeC}

The main process $e^-p \to e^-\gamma^*\gamma^*p \to e^-W^+W^-p$ is given in Fig. 1. In the case of investigating the anomalous $WW\gamma\gamma$ couplings, we take into account the sub-process $\gamma^*\gamma^* \to W^+W^-$ of the main process $e^-p \to e^-\gamma^*\gamma^*p \to e^-W^+W^-p$. The
anomalous $WW\gamma\gamma$ vertex contributions are shown in the first diagram of Fig. 2, whereas the others depict the SM Feynman diagrams.
Here, $\gamma^*$ flux in $\gamma^* \gamma^*$ collisions is defined by the Weizsacker-Williams approximation (WWA)\cite{Weizsacker,Williams}.
The WWA, which is also known as method of virtual quanta, is a semiclassical approximation. The idea of this approximation is that the electromagnetic field generated by a fast moving charged particle is nearly transverse which is like a plane wave and can be approximated by real photon.

\begin{figure}[h]
\centerline{\scalebox{0.6}{\includegraphics{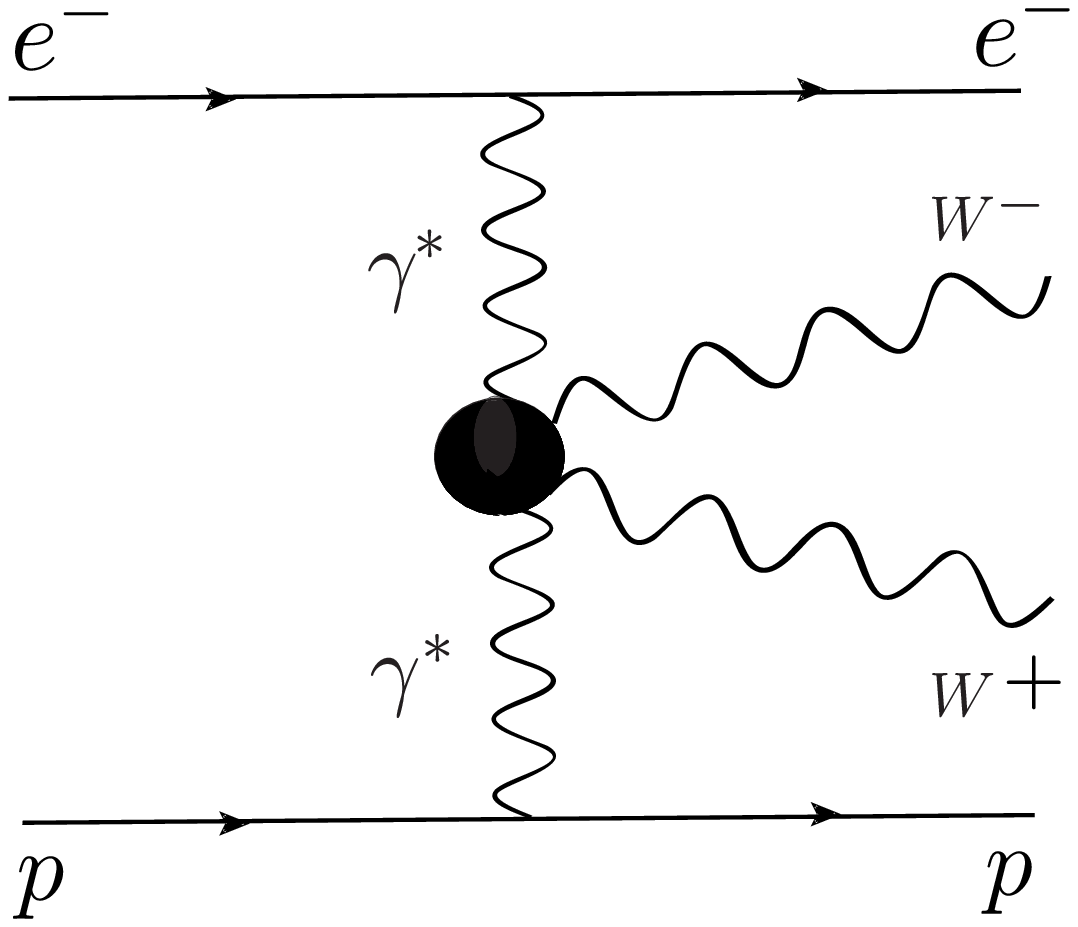}}}
\caption{ \label{fig:gamma1} A schematic diagram for the processes $e^-p \to e^-\gamma^*\gamma^*p \to e^-W^+W^-p$.}
\label{Fig.1}
\end{figure}

\begin{figure}[ht]
\centerline{\scalebox{0.7}{\includegraphics{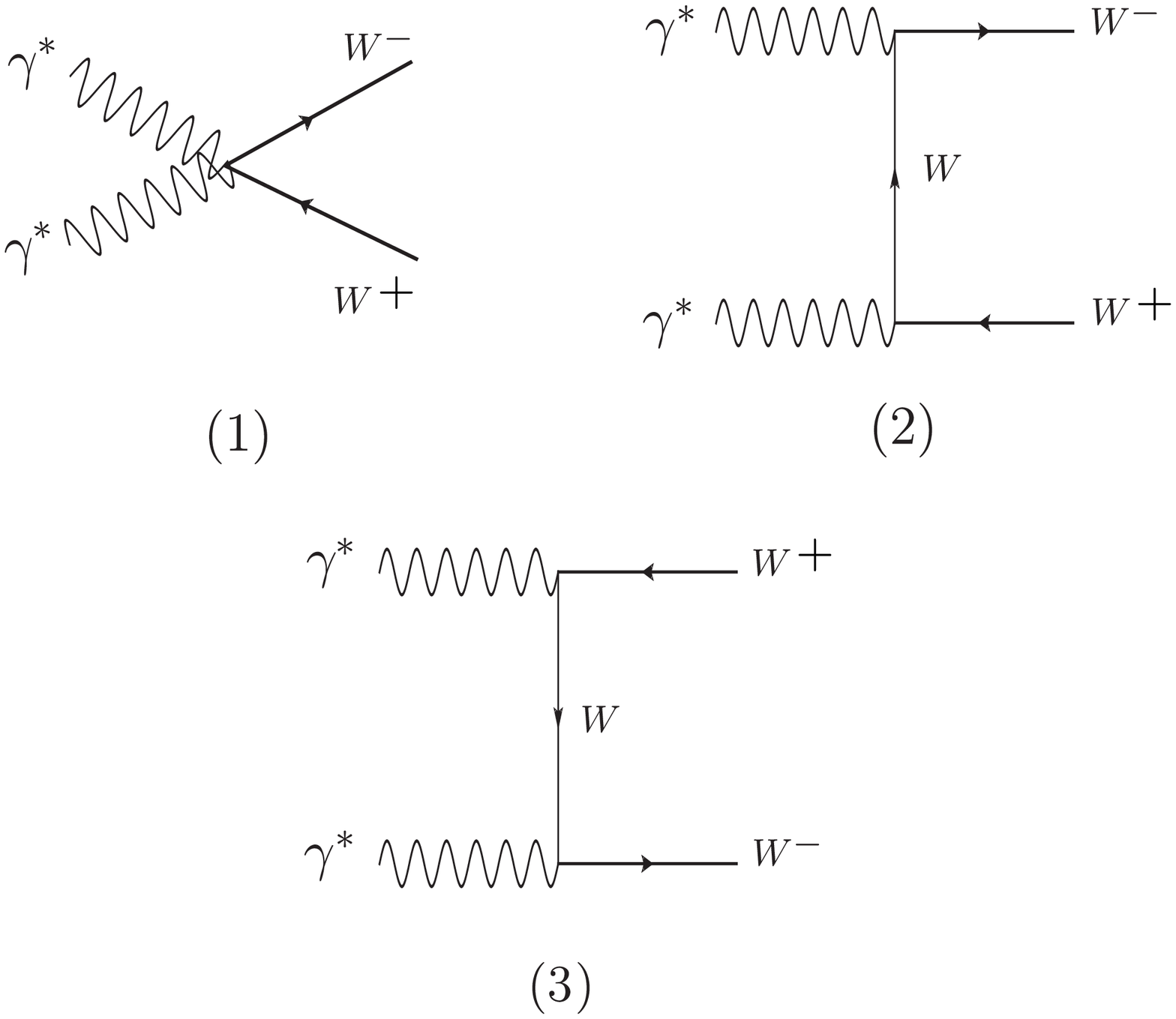}}}
\caption{ \label{fig:gamma2} Feynman diagrams contributing to the subprocess $\gamma^*\gamma^* \to W^+W^-$.}
\label{Fig.2}
\end{figure}

In the examined process, the effective Lagrangians with the anomalous quartic couplings are implemented to FeynRules package \cite{rul} and embedded into MadGraph$5_{-}$aMC$@$NLO \cite{mad} as a Universal FeynRules Output \cite{ufo}. In order to examine the possibilities of the LHeC and FCC-he as an option to sensitivity estimates on the anomalous quartic $WW \gamma \gamma$ couplings, we focus on the $e^-p \to e^-\gamma^*\gamma^*p \to e^-W^+W^-p$ signal and background process. Here, we choose the following set of basic cuts in the process $ep\to\nu_{e}W^{+}W^{-}$j containing the anomalous quartic $WW\gamma \gamma$ vertex. For pure leptonic decay channel, these cuts are given by

\begin{eqnarray}
p_{T_{j}}>20 GeV,p_{T_{\ell}}>10 GeV,
\end{eqnarray}

\begin{eqnarray}
 |\eta_{j}|< 5,|\eta_{\ell}|< 2.5,
\end{eqnarray}

\begin{eqnarray}
\Delta R(l,l)>0.4,\Delta R(j,l)>0.4
\end{eqnarray}

for semileptonic decay channel, applied cuts are

\begin{eqnarray}
p_{T_{j}}>20 GeV,p_{T_{\ell}}>10 GeV,
\end{eqnarray}

\begin{eqnarray}
 |\eta_{j}|< 5,|\eta_{\ell}|< 2.5,
\end{eqnarray}

\begin{eqnarray}
\Delta R(j,l)>0.4,\Delta R(j,j)>0.4,
\end{eqnarray}
where $\eta$ is the pseudorapidity, $p_T$ and $\Delta R$ are the transverse momentum and the separation of the final state particles, respectively.

In the WWA, two photons are used in the subprocess $\gamma^{*} \gamma^{*} \rightarrow W^{+} W^{-}$. The spectrum of first photon emitted by electron is given as \cite{Budnev,Chen2}

\begin{eqnarray}
f_{\gamma^{*}_{1}}(x_{1})=\frac{\alpha}{\pi E_{e}}\{[\frac{1-x_{1}+x_{1}^{2}/2}{x_{1}}]log(\frac{Q_{max}^{2}}{Q_{min}^{2}})-\frac{m_{e}^{2}x_{1}}{Q_{min}^{2}}
&&(1-\frac{Q_{min}^{2}}{Q_{max}^{2}})-\frac{1}{x_{1}}[1-\frac{x_{1}}{2}]^{2}log(\frac{x_{1}^{2}E_{e}^{2}+Q_{max}^{2}}{x_{1}^{2}E_{e}^{2}+Q_{min}^{2}})\} \nonumber \\
\end{eqnarray}
where $x_{1}=E_{\gamma_{1}^{*}}/E_{e}$ and $Q_{max}^{2}$ is maximum virtuality of the photon. The minimum value of $Q_{min}^{2}$ is shown as follows

\begin{eqnarray}
Q_{min}^{2}=\frac{m_{e}^{2}x_{1}^{2}}{1-x_{1}}.
\end{eqnarray}

Second, the spectrum of second photon emitted by proton can be written as follows \cite{Budnev,Chen2}

\begin{eqnarray}
f_{\gamma^{*}_{2}}(x_{2})=\frac{\alpha}{\pi E_{p}}\{[1-x_{2}][\varphi(\frac{Q_{max}^{2}}{Q_{0}^{2}})-\varphi(\frac{Q_{min}^{2}}{Q_{0}^{2}})]
\end{eqnarray}

where the function $\varphi$ is given by

\begin{eqnarray}
\varphi(\theta)=&&(1+ay)\left[-\textit{In}(1+\frac{1}{\theta})+\sum_{k=1}^{3}\frac{1}{k(1+\theta)^{k}}\right]+\frac{y(1-b)}{4\theta(1+\theta)^{3}} \nonumber \\
&& +c(1+\frac{y}{4})\left[\textit{In}\left(\frac{1-b+\theta}{1+\theta}\right)+\sum_{k=1}^{3}\frac{b^{k}}{k(1+\theta)^{k}}\right]. \nonumber \\
\end{eqnarray}
Here,

\begin{eqnarray}
y=\frac{x_{2}^{2}}{(1-x_{2})},
\end{eqnarray}
\begin{eqnarray}
a=\frac{1+\mu_{p}^{2}}{4}+\frac{4m_{p}^{2}}{Q_{0}^{2}}\approx 7.16,
\end{eqnarray}
\begin{eqnarray}
b=1-\frac{4m_{p}^{2}}{Q_{0}^{2}}\approx -3.96,
\end{eqnarray}
\begin{eqnarray}
c=\frac{\mu_{p}^{2}-1}{b^{4}}\approx 0.028.
\end{eqnarray}

Therefore, with this methodology, the total cross-section of the $e^-p \to e^-\gamma^*\gamma^*p \to e^-W^+W^-p$ process
at the LHeC and the FCC-he is obtained from:

\begin{eqnarray}
\sigma=\int f_{\gamma^*}(x_1)f_{\gamma^*}(x_2) d\hat{\sigma}_{\gamma^*\gamma^*}dE_{1} dE_{2}.
\end{eqnarray}

We calculate the dependencies of the $e^-p \to e^-\gamma^*\gamma^*p \to e^-W^+W^-p$ process production cross-sections
$\sigma(f_{M,i}, f_{T,i}, \sqrt{s})$ on $f_{M,i}$ and $f_{T,i}$ for the LHeC at $\sqrt{s}=1.30$
and 1.98 TeV. In Figs. 3-4 and 5-6, our numerical results for the total cross-section as a function of the aQGC $f_{M,i}$ ($f_{T,i}$) are summarized.
We consider center-of-mass energies $\sqrt{s}=1.30, 1.98$ TeV and we focus on the pure-leptonic and semi-leptonic decay channels for the
$W^+W^-$ bosons. It is observed from these figures that the cross-section is sensitive to the anomalous couplings $f_{M,i}$ and $f_{T,i}$.
In addition, the cross-section increases as $f_{M,i}$ ($f_{T,i}$) increase. For example, we can see that $\sigma(f_ {T, 5}, \sqrt{s}) \gg
\sigma(f_ {M, 7}, \sqrt{s})$ for several orders of magnitude, which implies that the obtained limits on $\sigma(f_ {T, 5}, \sqrt{s})$ are much more sensitive
than with respect to $f_{M,7}$, as well as with the other $f_ {M, i}$ and $f_ {T, i}$ parameters. This is a result of the energy dependence of the dimension-8 operators.  It is appropriate to mention that the differences observed in Figs. 3-6 with respect to the aQGC $f_{M,i}$ ($f_{T,i}$) can also be seen in Tables II and III, respectively.

\begin{figure}[h]
\centerline{\scalebox{1.3}{\includegraphics{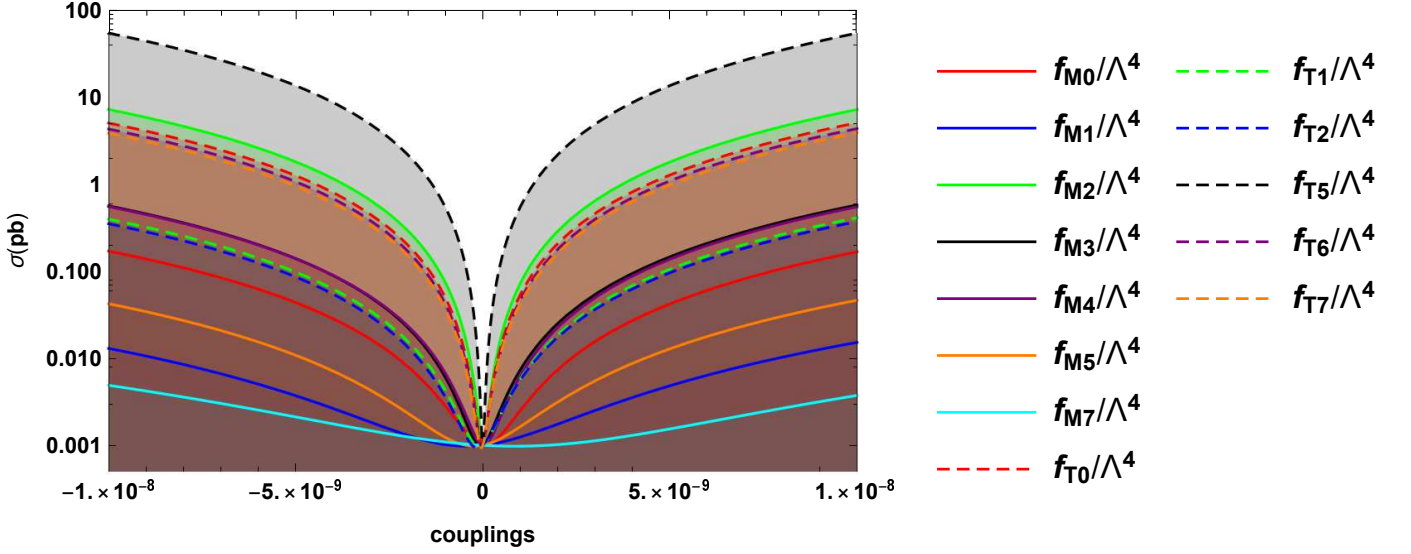}}}
\caption{ \label{fig:gamma1} For pure-leptonic channel, the total cross-sections of the process $e^-p \to e^-\gamma^*\gamma^*p \to e^-W^+W^-p$
as a function of the anomalous couplings for center-of-mass energy $\sqrt{s}=1.30$ TeV at the LHeC.}
\label{Fig.1}
\end{figure}

\begin{figure}[h]
\centerline{\scalebox{1.3}{\includegraphics{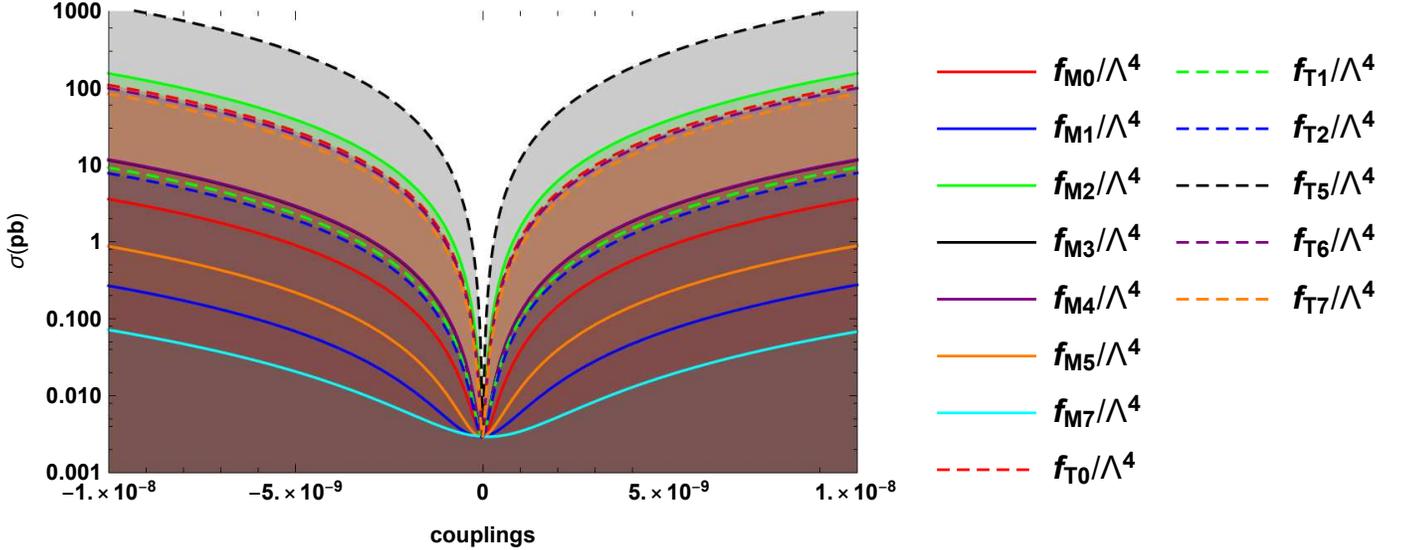}}}
\caption{ \label{fig:gamma2} Same as in Fig. 3, but for $\sqrt{s}=1.98$ TeV at the LHeC.}
\label{Fig.2}
\end{figure}

\begin{figure}[h]
\centerline{\scalebox{1.3}{\includegraphics{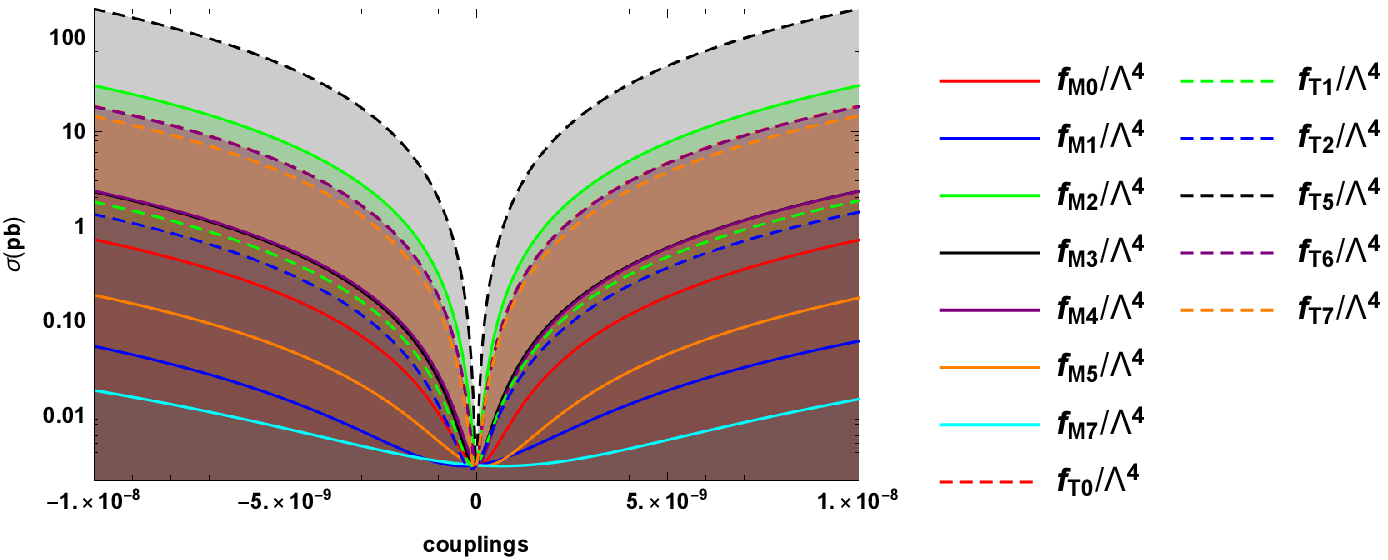}}}
\caption{ Same as in Fig. 3, but for semi-leptonic decay.}
\label{Fig.3}
\end{figure}

\begin{figure}[h]
\centerline{\scalebox{1.3}{\includegraphics{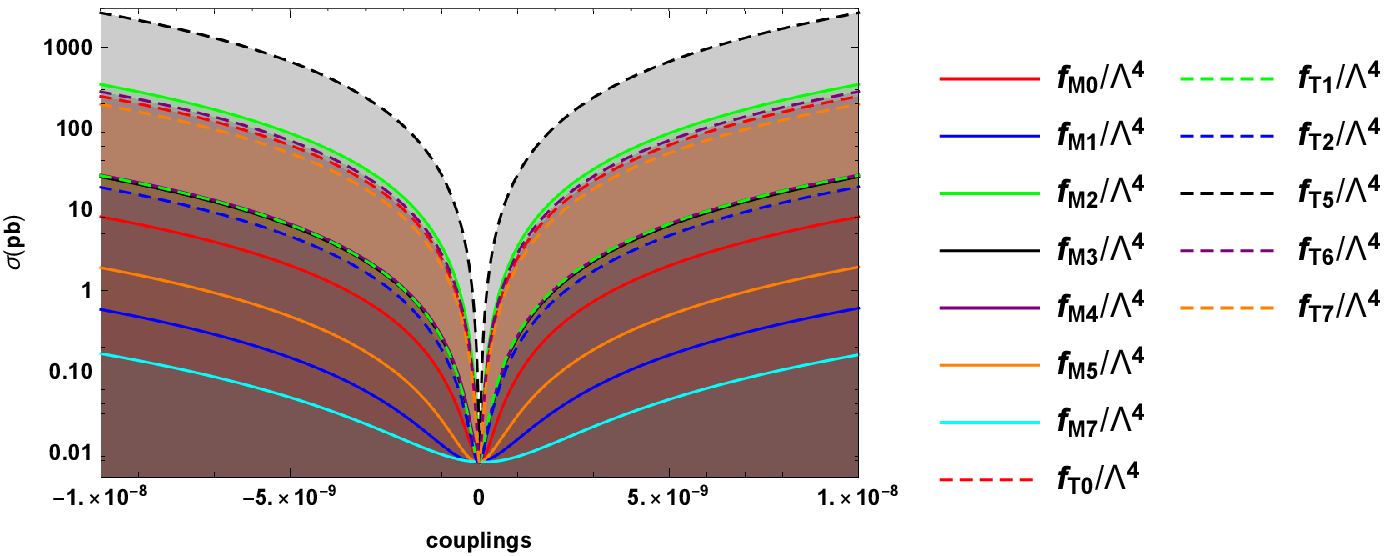}}}
\caption{ Same as in Fig. 4, but for semi-leptonic decay.}
\label{Fig.4}
\end{figure}

\begin{table} [h]
\caption{The total cross-sections of the $e^-p \to e^-\gamma^*\gamma^*p \to e^-W^+W^-p$ signal for $\sqrt{s}=1.30, 1.98$ TeV
at the LHeC depending on the anomalous couplings obtained by dimension-8 operators. The total cross-sections for each
coupling are calculated while fixing the other couplings to zero. The pure-leptonic decay channel of the $W^+W^-$ in the final state are considered.}
\begin{tabular}{|c|c|c|}
\hline\hline
Couplings (TeV$^{-4}$)  & $\sigma_{ep} (\mbox{pb})@ \sqrt{s}=1.30$ TeV &  $\sigma_{ep}(\mbox{pb})@ \sqrt{s}=1.98$ TeV    \\
\hline \hline
$f_{M0}/\Lambda^{4}$  & 1.67 $\times 10^{-1}$ & 3.57 \\
\hline
$f_{M1}/\Lambda^{4}$  & 1.54 $\times 10^{-2}$ & 2.76 $ \times 10^{-1}$ \\
\hline
$f_{M2}/\Lambda^{4}$  & 7.22  & 1.54 $\times 10^{2}$  \\
\hline
$f_{M3}/\Lambda^{4}$  & 5.81 $\times 10^{-1}$  & 1.15 $\times 10^{1}$ \\
\hline
$f_{M4}/\Lambda^{4}$  & 5.50 $\times 10^{-1}$  & 1.17 $\times 10^{1}$ \\
\hline
$f_{M5}/\Lambda^{4}$  & 4.22 $\times 10^{-2}$  & 8.71 $\times 10^{-1}$ \\
\hline
$f_{M7}/\Lambda^{4}$  & 3.76 $\times 10^{-3}$  & 6.81 $\times 10^{-2}$ \\
\hline
$f_{T0}/\Lambda^{4}$  &  5.13  & 1.09 $\times 10^{2}$   \\
\hline
$f_{T1}/\Lambda^{4}$  &  4.14 $\times 10^{-1}$ & 9.18  \\
\hline
$f_{T2}/\Lambda^{4}$  &  3.73 $\times 10^{-1}$ & 7.84  \\
\hline
$f_{T5}/\Lambda^{4}$  & 5.49 $\times 10^{1}$  & 1.17 $\times 10^{3}$  \\
\hline
$f_{T6}/\Lambda^{4}$  & 4.38  & 9.85 $\times 10^{1}$  \\
\hline
$f_{T7}/\Lambda^{4}$  & 3.92  & 8.40 $\times 10^{1}$  \\
\hline
\end{tabular}
\end{table}

\begin{table} [h]
\caption{ The total cross-sections of the $e^-p \to e^-\gamma^*\gamma^*p \to e^-W^+W^-p$ signal for $\sqrt{s}=1.30, 1.98$ TeV
at the LHeC depending on the anomalous couplings obtained by dimension-8 operators. The total cross-sections for each
coupling are calculated while fixing the other couplings to zero. The semi-leptonic decay channel of the $W^+W^-$ in the final state are considered.}
\begin{tabular}{|c|c|c|}
\hline\hline
Couplings (TeV$^{-4}$)  & $\sigma_{ep} (\mbox{pb})@ \sqrt{s}=1.30$ TeV &  $\sigma_{ep}(\mbox{pb})@ \sqrt{s}=1.98$ TeV    \\
\hline \hline
$f_{M0}/\Lambda^{4}$  & 7.07 $\times 10^{-1}$ & 8.04\\
\hline
$f_{M1}/\Lambda^{4}$  & 6.04 $\times 10^{-2}$ & 6.06 $ \times 10^{-1}$ \\
\hline
$f_{M2}/\Lambda^{4}$  & 3.05 $\times 10^{1}$  & 3.46 $\times 10^{2}$ \\
\hline
$f_{M3}/\Lambda^{4}$  & 2.33  & 2.52 $\times 10^{1}$ \\
\hline
$f_{M4}/\Lambda^{4}$  & 2.32  & 2.64 $\times 10^{1}$ \\
\hline
$f_{M5}/\Lambda^{4}$  & 1.73 $\times 10^{-1}$  & 1.90 \\
\hline
$f_{M7}/\Lambda^{4}$  & 1.46 $\times 10^{-2}$  & 1.49 $\times 10^{-1}$ \\
\hline
$f_{T0}/\Lambda^{4}$  &  1.86 $\times 10^{1}$ & 2.47 $\times 10^{2}$ \\
\hline
$f_{T1}/\Lambda^{4}$  &  1.85  & 2.64 $\times 10^{1}$  \\
\hline
$f_{T2}/\Lambda^{4}$  &  1.40 & 1.86 $\times 10^{1}$ \\
\hline
$f_{T5}/\Lambda^{4}$  & 1.99 $\times 10^{2}$ & 2.64 $\times 10^{3}$   \\
\hline
$f_{T6}/\Lambda^{4}$  & 1.96 $\times 10^{1}$  & 2.81 $\times 10^{2}$  \\
\hline
$f_{T7}/\Lambda^{4}$  & 1.48 $\times 10^{1}$  & 1.99 $\times 10^{2}$ \\
\hline
\end{tabular}
\end{table}

To complement our study, specifically our results at $95\%$ C. L. on the aQGC $f_{M,i}$ and $f_{T,i}$ are obtained using
$\chi^2$ analysis \cite{Koksal6,Gutierrez2,Billur5,Koksal7,Gutierrez3} with $\sqrt{s}=1.30, 1.98$ TeV and ${\cal L}=10, 30, 50, 100$ ${\rm fb^{-1}}$

\begin{equation}
\chi^2(f_{M,i}, f_{T,i})=\Biggl(\frac{\sigma_{SM}-\sigma_{BSM}(\sqrt{s}, f_{M,i}, f_{T,i})}
{\sigma_{SM}\sqrt{(\delta_{st})^2+(\delta_{sys})^2}}\Biggr)^2
\end{equation}

\noindent with $\sigma_{BSM}(\sqrt{s}, f_{M,i}, f_{T,i})$ and $\sigma_{SM}$ are the cross-sections in the presence of BSM interactions
and in the SM, respectively. $\delta_{st}=\frac{1}{\sqrt{N_{SM}}}$ is the statistical error and $\delta_{sys}$ is the systematic error.
The number of events is given by $N_{SM}={\cal L}_{int}\times \sigma_{SM}$, where ${\cal L}_{int}$ is the
integrated luminosity of the LHeC.

Tables IV-VII summarize the limits on the dimension-8 aQGC parameters $f_{M,i}/\Lambda^{4}$ and $f_{T,i}/\Lambda^{4}$
obtained from 1.30, 1.98 TeV and ${\cal L}=10, 30, 50, 100$ ${\rm fb^{-1}}$ data separately. Where both
pure-leptonic and semi-leptonic decay channels of the $W^+W^-$ bosons in the final state are considered. The most restrictive limits
at $95\%$ C.L. are obtained for $f_{T,5}/\Lambda^{4}$ followed by $f_{M,2}/\Lambda^{4}$, $f_{T,0}/\Lambda^{4}$, etc.. Comparing
our results with the corresponding experimental results reported in Table I, we conclude that our results are of the same order
of magnitude as the experimental results reported by the CMS Collaboration previously (see Table I) and compare favorably with
other results reported in the literature by various authors and in other contexts. The only work on the anomalous $WW\gamma \gamma$ coupling in future $ep$ colliders is examined by Ref. \cite{Ari1}. In that study, the anomalous $WW\gamma \gamma$ coupling is investigated through $e^-p \to \nu_e\gamma\gamma j$ at the LHeC and the FCC-he. The $e^-p \to \nu_e\gamma\gamma j$ process, which has been studied at the LHeC and the FCC-he, is particularly important because it allows us to make a direct comparison with our results for the aQGC, as well as with the experimental results reported by the CMS Collaboration (see Table I). In Ref. \cite{Ari1}, the limits obtained for the aQGC in the case of the FCC-he are up to two orders of magnitude better than those reported by the CMS Collaboration, and with respect to our results they are up to an order of magnitude better.

\begin{table} [h]
\caption{Limits at $95\%$ C.L. on the anomalous $WW\gamma\gamma$ quartic couplings of the $e^-p \to e^-\gamma^*\gamma^*p \to e^-W^+W^-p$
signal for $\sqrt{s}=1.30$ TeV at the LHeC. The coupling are calculated while fixing the other couplings to zero.
The pure-leptonic decay channel of the $W^+W^-$ in the final state are considered.}
\begin{tabular}{|c|c|c|c|c|c|}
\hline\hline
Couplings (TeV$^{-4}$) & 10 fb$^{-1}$ & 30 fb$^{-1}$ & 50 fb$^{-1}$ & 100 fb$^{-1}$ \\
\hline \hline
$f_{M0}/\Lambda^{4}$  & [-0.56; 0.66] $\times 10^{3}$ & [-0.42; 0.51] $\times 10^{3}$ & [-0.36; 0.46] $\times 10^{3}$ & [-0.30; 0.39] $\times 10^{3}$ \\
\hline
$f_{M1}/\Lambda^{4}$  & [-0.26; 0.18] $\times 10^{4}$ & [-0.21; 0.13]$ \times 10^{4}$  & [-0.19; 0.11] $ \times 10^{4}$ & [-0.17; 0.09] $ \times 10^{4}$   \\
\hline
$f_{M2}/\Lambda^{4}$  & [-0.87; 0.99] $\times 10^{2}$ & [-0.65; 0.76]$ \times 10^{2}$  & [-0.57; 0.68] $ \times 10^{2}$ & [-0.47; 0.58]$\times 10^{2}$   \\
\hline
$f_{M3}/\Lambda^{4}$  & [-0.39; 0.28] $\times 10^{3}$ & [-0.32; 0.20]$ \times 10^{3}$  & [-0.29; 0.17] $ \times 10^{3}$ & [-0.25; 0.14]$\times 10^{3}$   \\
\hline
$f_{M4}/\Lambda^{4}$  & [-0.32; 0.36] $\times 10^{3}$ & [-0.24;0.28]$ \times 10^{3}$  & [-0.21; 0.25] $ \times 10^{3}$ & [-0.17; 0.21]$\times 10^{3}$   \\
\hline
$f_{M5}/\Lambda^{4}$  & [-1.45; 0.99] $\times 10^{3}$ & [-1.17; 0.70]$ \times 10^{3}$  & [-1.07; 0.60] $ \times 10^{3}$ & [-0.95; 0.48]$\times 10^{3}$   \\
\hline
$f_{M7}/\Lambda^{4}$  & [-0.36; 0.53]$\times 10^{4}$    & [-0.25; 0.43]$\times 10^{4}$   & [-0.22; 0.39]$\times 10^{4}$   & [-0.17; 0.35]$\times 10^{4}$    \\
\hline
$f_{T0}/\Lambda^{4}$  &  [-1.33; 0.93] $\times 10^{2}$ & [-1.07; 0.67] $\times 10^{2}$ & [-0.97; 0.57] $\times 10^{2}$ & [-0.86; 0.45] $\times 10^{2}$ \\
\hline
$f_{T1}/\Lambda^{4}$  & [-0.50; 0.31] $\times 10^{3}$  & [-0.41; 0.22] $\times 10^{3}$ & [-0.37; 0.19] $\times 10^{3}$ & [-0.33; 0.15] $\times 10^{3}$  \\
\hline
$f_{T2}/\Lambda^{4}$  &  [-0.58; 0.30] $\times 10^{3}$ & [-0.49; 0.20] $\times 10^{3}$ & [-0.45; 0.17] $\times 10^{3}$ & [-0.42; 0.13] $\times 10^{3}$ \\
\hline
$f_{T5}/\Lambda^{4}$  & [-0.36; 0.32] $\times 10^{2}$  & [-0.28; 0.23] $\times 10^{2}$ & [-0.25; 0.20] $\times 10^{2}$ & [-0.22; 0.17] $\times 10^{2}$  \\
\hline
$f_{T6}/\Lambda^{4}$  & [-1.51; 0.95] $\times 10^{2}$  & [-1.23; 0.67] $\times 10^{2}$ & [-1.13; 0.57] $\times 10^{2}$ & [-1.01; 0.45] $\times 10^{2}$ \\
\hline
$f_{T7}/\Lambda^{4}$  & [-1.73; 0.93] $\times 10^{2}$  & [-1.45; 0.64] $\times 10^{2}$ & [-1.34; 0.54] $\times 10^{2}$ & [-1.22; 0.42] $\times 10^{2}$ \\
\hline
\end{tabular}
\end{table}

\begin{table} [h]
\caption{Limits at $95\%$ C.L. on the anomalous $WW\gamma\gamma$ quartic couplings of the $e^-p \to e^-\gamma^*\gamma^*p \to e^-W^+W^-p$
signal for $\sqrt{s}=1.98$ TeV at the LHeC. The coupling are calculated while fixing the other couplings to zero.
The pure-leptonic decay channel of the $W^+W^-$ in the final state are considered.}
\begin{tabular}{|c|c|c|c|c|c|}
\hline\hline
Couplings (TeV$^{-4}$) & 10 fb$^{-1}$ & 30 fb$^{-1}$ & 50 fb$^{-1}$ & 100 fb$^{-1}$ \\
\hline \hline
$f_{M0}/\Lambda^{4}$  & [-0.17; 0.18] $\times 10^{3}$ & [-0.12; 0.14] $\times 10^{3}$ & [-0.11; 0.12] $\times 10^{3}$ & [-0.09; 0.10] $\times 10^{3}$ \\
\hline
$f_{M1}/\Lambda^{4}$  & [-0.70; 0.57] $\times 10^{3}$ & [-0.55; 0.42]$ \times 10^{3}$  & [-0.50; 0.36] $ \times 10^{3}$ & [-0.43; 0.29] $ \times 10^{3}$   \\
\hline
$f_{M2}/\Lambda^{4}$  & [-0.25; 0.27] $\times 10^{2}$ & [-0.19; 0.21]$ \times 10^{2}$  & [-0.17; 0.19] $ \times 10^{2}$ & [-0.14; 0.16]$\times 10^{2}$   \\
\hline
$f_{M3}/\Lambda^{4}$  & [-1.01; 0.92] $\times 10^{2}$ & [-0.78; 0.69]$ \times 10^{2}$  & [-0.69; 0.60] $ \times 10^{2}$ & [-0.59; 0.50]$\times 10^{2}$   \\
\hline
$f_{M4}/\Lambda^{4}$  & [-0.95; 0.96] $\times 10^{2}$ & [-0.72; 0.73]$ \times 10^{2}$  & [-0.64; 0.65] $ \times 10^{2}$ & [-0.53; 0.54]$\times 10^{2}$   \\
\hline
$f_{M5}/\Lambda^{4}$  & [-0.39; 0.31] $\times 10^{3}$ & [-0.31; 0.23]$ \times 10^{3}$  & [-0.28; 0.20] $ \times 10^{3}$ & [-0.24; 0.16]$\times 10^{3}$   \\
\hline
$f_{M7}/\Lambda^{4}$  & [-0.11; 0.14]$\times 10^{4}$    & [-0.08; 0.11]$\times 10^{4}$   & [-0.72; 0.99]$\times 10^{3}$   & [-0.59;0.86]$\times 10^{3}$    \\
\hline
$f_{T0}/\Lambda^{4}$  &  [-0.36; 0.28] $\times 10^{2}$ & [-0.28; 0.20] $\times 10^{2}$ & [-0.25; 0.17] $\times 10^{2}$ & [-0.22; 0.14] $\times 10^{2}$ \\
\hline
$f_{T1}/\Lambda^{4}$  & [-1.27; 0.91] $\times 10^{2}$  & [-1.02; 0.66] $\times 10^{2}$ & [-0.92; 0.56] $\times 10^{2}$ & [-0.81; 0.45] $\times 10^{2}$  \\
\hline
$f_{T2}/\Lambda^{4}$  &  [-1.48; 0.93] $\times 10^{2}$ & [-1.21; 0.66] $\times 10^{2}$ & [-1.10; 0.56] $\times 10^{2}$ & [-0.99; 0.44] $\times 10^{2}$ \\
\hline
$f_{T5}/\Lambda^{4}$  & [-0.99; 0.93] $\times 10^{1}$  & [-0.76; 0.70] $\times 10^{1}$ & [-0.68; 0.61] $\times 10^{1}$ & [-0.57; 0.51] $\times 10^{1}$  \\
\hline
$f_{T6}/\Lambda^{4}$  & [-0.40; 0.27] $\times 10^{2}$  & [-0.32; 0.20] $\times 10^{2}$ & [-0.29; 0.17] $\times 10^{2}$ & [-0.26; 0.13] $\times 10^{2}$ \\
\hline
$f_{T7}/\Lambda^{4}$  & [-0.45; 0.28] $\times 10^{2}$  & [-0.39; 0.19] $\times 10^{2}$ & [-0.36; 0.16] $\times 10^{2}$ & [-0.32; 0.12] $\times 10^{2}$ \\
\hline
\end{tabular}
\end{table}

\begin{table} [h]
\caption{Limits at $95\%$ C.L. on the anomalous $WW\gamma\gamma$ quartic couplings of the $e^-p \to e^-\gamma^*\gamma^*p \to e^-W^+W^-p$
signal for $\sqrt{s}=1.30$ TeV at the LHeC. The coupling are calculated while fixing the other couplings to zero.
The semi-leptonic decay channel of the $W^+W^-$ in the final state are considered.}
\begin{tabular}{|c|c|c|c|c|c|}
\hline\hline
Couplings (TeV$^{-4}$) & 10 fb$^{-1}$ & 30 fb$^{-1}$ & 50 fb$^{-1}$ & 100 fb$^{-1}$ \\
\hline \hline
$f_{M0}/\Lambda^{4}$  & [-0.36; 0.42] $\times 10^{3}$ & [-0.27; 0.33] $\times 10^{3}$ & [-0.23; 0.29] $\times 10^{3}$ & [-0.19; 0.25] $\times 10^{3}$ \\
\hline
$f_{M1}/\Lambda^{4}$  & [-0.18; 0.11] $\times 10^{4}$ & [-0.15; 0.08]$ \times 10^{4}$  & [-0.13; 0.07] $ \times 10^{4}$ & [-0.12; 0.05] $ \times 10^{4}$   \\
\hline
$f_{M2}/\Lambda^{4}$  & [-0.58; 0.59] $\times 10^{2}$ & [-0.44; 0.45]$ \times 10^{2}$  & [-0.39; 0.40] $ \times 10^{2}$ & [-0.33; 0.34]$\times 10^{2}$   \\
\hline
$f_{M3}/\Lambda^{4}$  & [-0.27; 0.18] $\times 10^{3}$ & [-0.22; 0.13]$ \times 10^{3}$  & [-0.20; 0.11] $ \times 10^{3}$ & [-0.17; 0.08]$\times 10^{3}$   \\
\hline
$f_{M4}/\Lambda^{4}$  & [-0.20; 0.23] $\times 10^{3}$ & [-0.15; 0.18]$ \times 10^{3}$  & [-0.13; 0.16] $ \times 10^{3}$ & [-0.10; 0.14]$\times 10^{3}$   \\
\hline
$f_{M5}/\Lambda^{4}$  & [-0.63; 0.97] $\times 10^{3}$ & [-0.45; 0.79]$ \times 10^{3}$  & [-0.38; 0.72] $ \times 10^{3}$ & [-0.30; 0.64]$\times 10^{3}$   \\
\hline
$f_{M7}/\Lambda^{4}$  & [-0.22; 0.35]$\times 10^{4}$    & [-0.16; 0.29]$\times 10^{4}$   & [-0.13; 0.26]$\times 10^{4}$   & [-0.11; 0.24]$\times 10^{4}$    \\
\hline
$f_{T0}/\Lambda^{4}$  &  [-0.95; 0.61] $\times 10^{2}$ & [-0.77; 0.43] $\times 10^{2}$ & [-0.70; 0.37] $\times 10^{2}$ & [-0.63; 0.29] $\times 10^{2}$ \\
\hline
$f_{T1}/\Lambda^{4}$  & [-0.33;0.18] $\times 10^{3}$  & [-0.28; 0.12] $\times 10^{3}$ & [-0.26; 0.10] $\times 10^{3}$ & [-0.23; 0.08] $\times 10^{3}$  \\
\hline
$f_{T2}/\Lambda^{4}$  &  [-0.44; 0.18] $\times 10^{3}$ & [-0.38; 0.12] $\times 10^{3}$ & [-0.36; 0.10] $\times 10^{3}$ & [-0.34; 0.07] $\times 10^{3}$ \\
\hline
$f_{T5}/\Lambda^{4}$  & [-0.29; 0.18] $\times 10^{2}$  & [-0.24; 0.13] $\times 10^{2}$ & [-0.22; 0.11] $\times 10^{2}$ & [-0.20; 0.09] $\times 10^{2}$  \\
\hline
$f_{T6}/\Lambda^{4}$  & [-0.91; 0.64] $\times 10^{2}$  & [-0.73; 0.46] $\times 10^{2}$ & [-0.67; 0.39] $\times 10^{2}$ & [-0.59; 0.31] $\times 10^{2}$ \\
\hline
$f_{T7}/\Lambda^{4}$  & [-1.40; 0.52] $\times 10^{2}$  & [-1.22; 0.35] $\times 10^{2}$ & [-1.16; 0.28] $\times 10^{2}$ & [-1.09; 0.21] $\times 10^{2}$ \\
\hline
\end{tabular}
\end{table}

\begin{table} [h]
\caption{Limits at $95\%$ C.L. on the anomalous $WW\gamma\gamma$ quartic couplings of the $e^-p \to e^-\gamma^*\gamma^*p \to e^-W^+W^-p$
signal for $\sqrt{s}=1.98$ TeV at the LHeC. The coupling are calculated while fixing the other couplings to zero.
The semi-leptonic decay channel of the $W^+W^-$ in the final state are considered.}
\begin{tabular}{|c|c|c|c|c|c|}
\hline\hline
Couplings (TeV$^{-4}$) & 10 fb$^{-1}$ & 30 fb$^{-1}$ & 50 fb$^{-1}$ & 100 fb$^{-1}$ \\
\hline \hline
$f_{M0}/\Lambda^{4}$  & [-0.14; 0.16] $\times 10^{3}$ & [-0.10; 0.12] $\times 10^{3}$ & [-0.09; 0.11] $\times 10^{3}$ & [-0.07; 0.09] $\times 10^{3}$ \\
\hline
$f_{M1}/\Lambda^{4}$  & [-0.64; 0.46] $\times 10^{3}$ & [-0.51; 0.33]$ \times 10^{3}$  & [-0.46; 0.28] $ \times 10^{3}$ & [-0.41; 0.23] $ \times 10^{3}$   \\
\hline
$f_{M2}/\Lambda^{4}$  & [-0.22; 0.23] $\times 10^{2}$ & [-0.16; 0.18]$ \times 10^{2}$  & [-0.14; 0.16] $ \times 10^{2}$ & [-0.12; 0.13]$\times 10^{2}$   \\
\hline
$f_{M3}/\Lambda^{4}$  & [-0.90; 0.76] $\times 10^{2}$ & [-0.71; 0.56]$ \times 10^{2}$  & [-0.63; 0.49] $ \times 10^{2}$ & [-0.54; 0.40]$\times 10^{2}$   \\
\hline
$f_{M4}/\Lambda^{4}$  & [-0.80; 0.82] $\times 10^{2}$ & [-0.61; 0.62]$ \times 10^{2}$  & [-0.53; 0.55] $ \times 10^{2}$ & [-0.45; 0.46]$\times 10^{2}$   \\
\hline
$f_{M5}/\Lambda^{4}$  & [-0.35; 0.26] $\times 10^{3}$ & [-0.28; 0.19]$ \times 10^{3}$  & [-0.25; 0.16] $ \times 10^{3}$ & [-0.22; 0.13]$\times 10^{3}$   \\
\hline
$f_{M7}/\Lambda^{4}$  & [-0.10; 0.11]$\times 10^{4}$    & [-0.72; 0.89]$\times 10^{3}$   & [-0.62; 0.80]$\times 10^{3}$   & [-0.51; 0.68]$\times 10^{3}$    \\
\hline
$f_{T0}/\Lambda^{4}$  &  [-0.31; 0.23] $\times 10^{2}$ & [-0.24; 0.17] $\times 10^{2}$ & [-0.22; 0.14] $\times 10^{2}$ & [-0.19; 0.12] $\times 10^{2}$ \\
\hline
$f_{T1}/\Lambda^{4}$  & [-0.97; 0.66] $\times 10^{2}$  & [-0.78; 0.47] $\times 10^{2}$ & [-0.71; 0.40] $\times 10^{2}$ & [-0.63; 0.32] $\times 10^{2}$  \\
\hline
$f_{T2}/\Lambda^{4}$  &  [-1.25; 0.73] $\times 10^{2}$ & [-1.04; 0.51] $\times 10^{2}$ & [-0.96; 0.43] $\times 10^{2}$ & [-0.86; 0.34] $\times 10^{2}$ \\
\hline
$f_{T5}/\Lambda^{4}$  & [-0.96; 0.67] $\times 10^{1}$  & [-0.78; 0.48] $\times 10^{1}$ & [-0.70; 0.41] $\times 10^{1}$ & [-0.62; 0.33] $\times 10^{1}$  \\
\hline
$f_{T6}/\Lambda^{4}$  & [-0.29; 0.21] $\times 10^{2}$  & [-0.23; 0.15] $\times 10^{2}$ & [-0.21; 0.13] $\times 10^{2}$ & [-0.18; 0.11] $\times 10^{2}$ \\
\hline
$f_{T7}/\Lambda^{4}$  & [-0.35; 0.25] $\times 10^{2}$  & [-0.28; 0.18] $\times 10^{2}$ & [-0.26; 0.15] $\times 10^{2}$ & [-0.23; 0.12] $\times 10^{2}$ \\
\hline
\end{tabular}
\end{table}

\section{Cross-section of the process $e^-p \to e^-\gamma^*\gamma^*p \to e^-W^+W^-p$ and limits on the aQGC
at the FCC-he}

Starting from the methodology that we presented in the previous sections for the evaluation of the total cross-section for the double
production of $W^\pm$ bosons, i.e. the $e^-p \to e^-\gamma^*\gamma^*p \to e^-W^+W^-p$ process as well as of the evaluation of the corresponding bounds for the aQGC $f_{M,i}/\Lambda^{4}$ and $f_{T,i}/\Lambda^{4}$ applying a $\chi^2$ analysis and where we adopt the kinematic cuts given by Eqs. (31)-(36), we give
our numerical computation for the center-of-mass energies and luminosities of the FCC-he, that is $\sqrt{s}=3.46, 5.29$ TeV and ${\cal L}=100, 300, 500, 1000$ ${\rm fb^{-1}}$, respectively.

In Figs. 7-10, the effect of the aQGC $f_{M,i}/\Lambda^{4}$ and $f_{T,i}/\Lambda^{4}$ on the total cross-section of the process
$e^-p \to e^-\gamma^*\gamma^*p \to e^-W^+W^-p$ is shown, where we compare the $\sigma(\sqrt{s},f_{M,i}/\Lambda^{4}, f_{T,i}/\Lambda^{4})$
expected for the different $f_{M,i}/\Lambda^{4}$ and $f_{T,i}/\Lambda^{4}$ parameters. The comparison of Figs. 9 and 10 with Figs. 7
and 8 shows the impact that the increase in the energy of the center-of-mass of the collider, the incorporation of the semi-leptonic
channel of the $W^\pm$ and the aQGC $f_{M,i}/\Lambda^{4}$ and $f_{T,i}/\Lambda^{4}$ can have on the $e^-p \to e^-\gamma^*\gamma^*p \to e^-W^+W^-p$
signal. The corresponding increase on the $\sigma(\sqrt{s},f_{M,i}/\Lambda^{4}, f_{T,i}/\Lambda^{4})$ is approximately one order of magnitude
stronger than the results shown in Figs. 7 and 8.

\begin{figure}[h]
\centerline{\scalebox{1.3}{\includegraphics{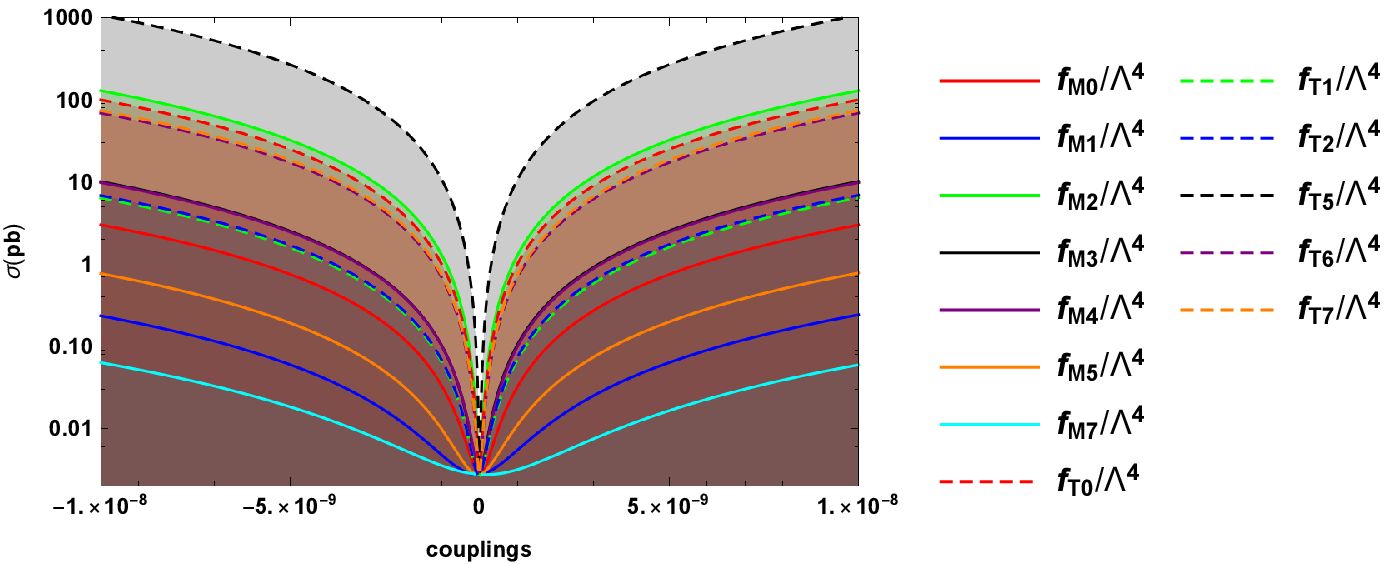}}}
\caption{ \label{fig:gamma1} For pure-leptonic channel, the total cross-sections of the process $e^-p \to e^-\gamma^*\gamma^*p \to e^-W^+W^-p$
as a function of the anomalous couplings for center-of-mass energy $\sqrt{s}=3.46$ TeV at the FCC-he.}
\label{Fig.1}
\end{figure}

\begin{figure}[h]
\centerline{\scalebox{1.3}{\includegraphics{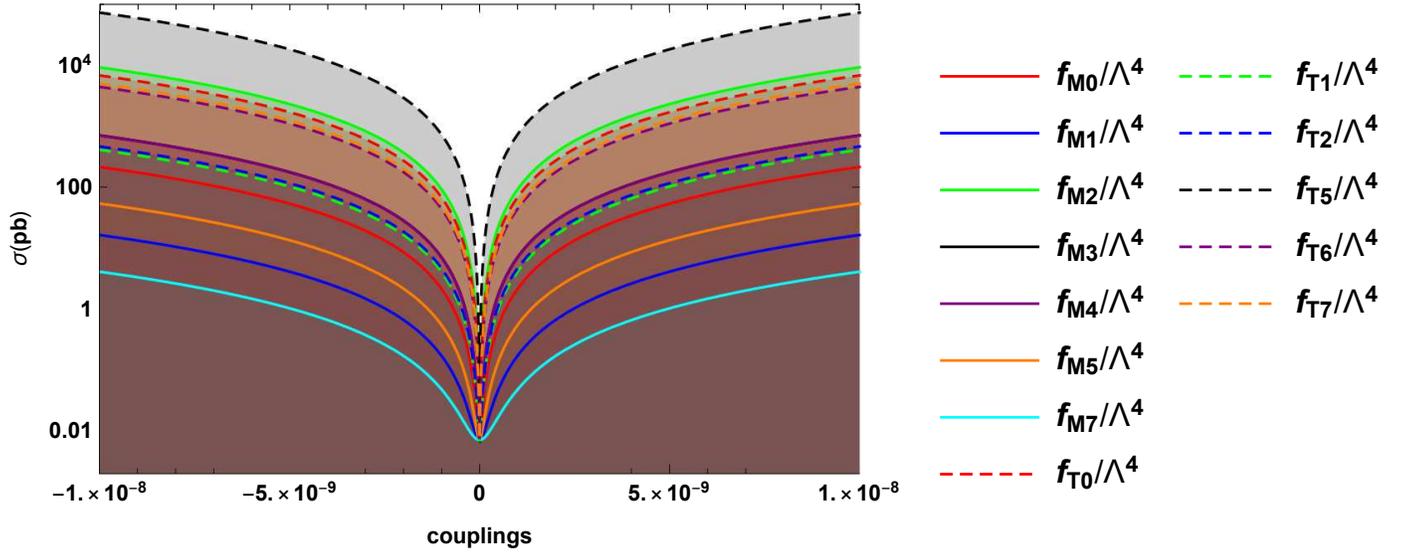}}}
\caption{ \label{fig:gamma2} Same as in Fig. 7, but for $\sqrt{s}=5.29$ TeV at the FCC-he.}
\label{Fig.2}
\end{figure}

\begin{figure}[h]
\centerline{\scalebox{1.3}{\includegraphics{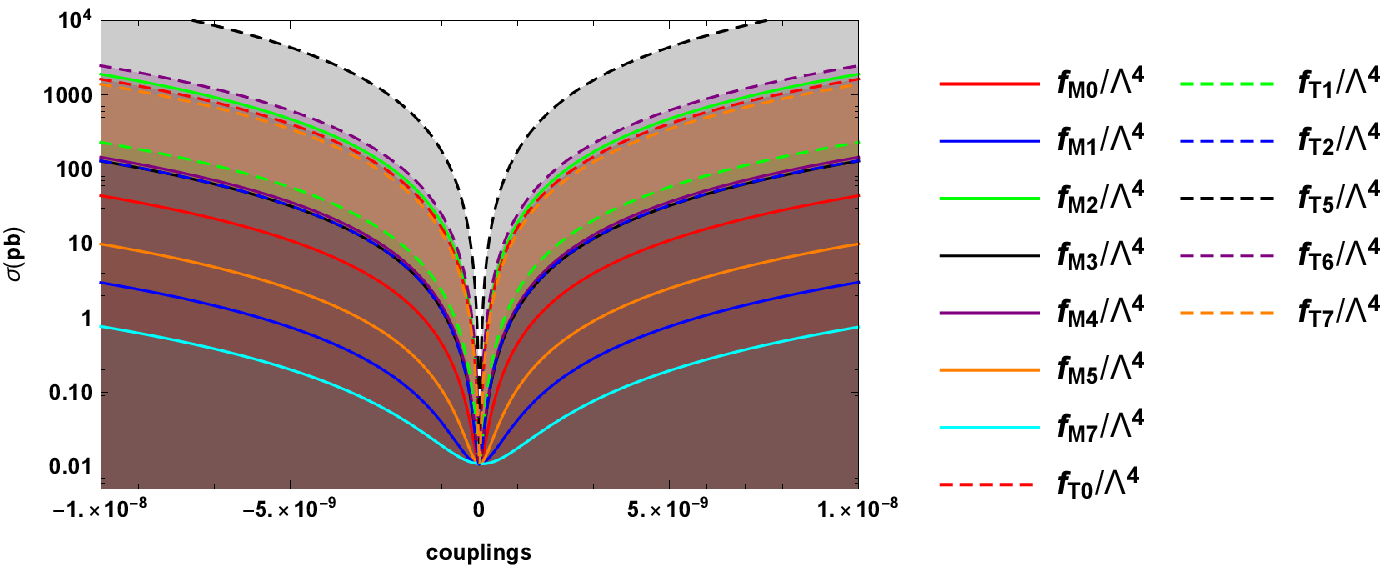}}}
\caption{ Same as in Fig. 7, but for semi-leptonic decay.}
\label{Fig.3}
\end{figure}

\begin{figure}[h]
\centerline{\scalebox{1.3}{\includegraphics{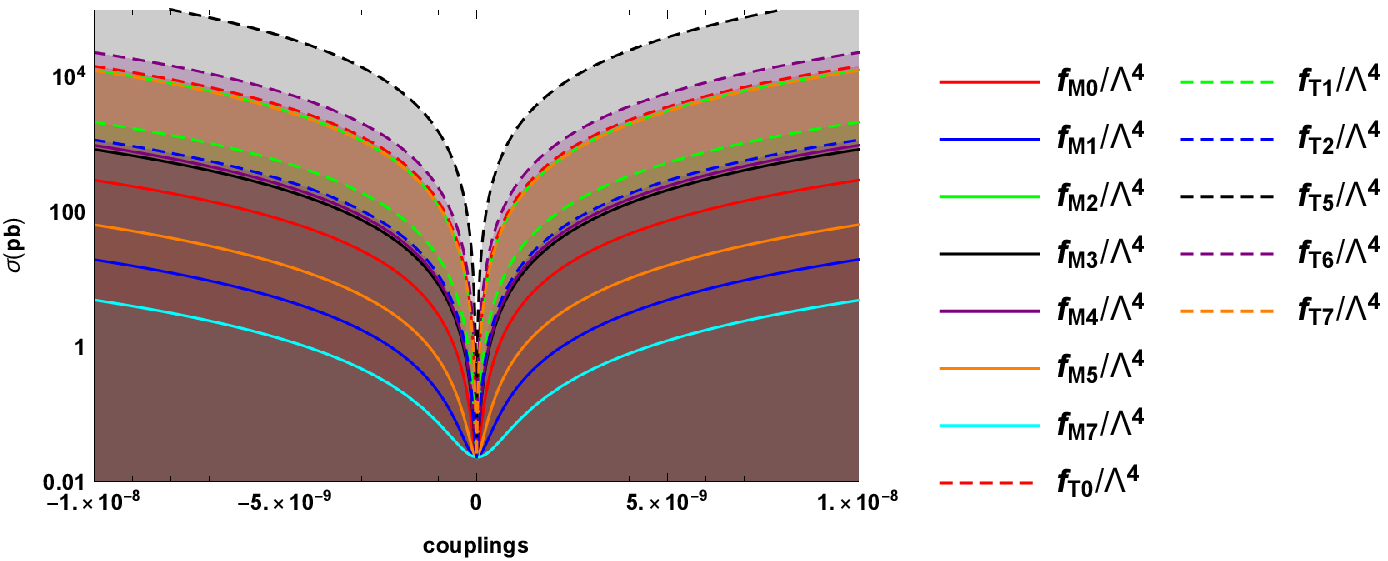}}}
\caption{ Same as in Fig. 8, but for semi-leptonic decay.}
\label{Fig.4}
\end{figure}

Tables VIII and IX show the measurements of the scattering cross-section of the $e^-p \to e^-\gamma^*\gamma^*p \to e^-W^+W^-p$ signal,
which is dependent on the aQGC $f_{M,i}/\Lambda^{4}$ and $f_{T,i}/\Lambda^{4}$. In this tables, the total cross-section for each coupling
are calculated while fixing the other couplings to zero. In Table VIII, the pure-leptonic decay channel of the $W^+W^-$ in the final
state is considered with $\sqrt{s}=3.46$ and 5.29 TeV at the FCC-he. However, in Table IX, the semi-leptonic decay channel of the $W^\pm$
is considered. The scattering cross-section enhancement for all the dimension-8 aQGC parameters is evident, but in the particular case
of the $f_{T,5}/\Lambda^{4}$ operator the impact is remarkable, and this behavior manifests for both center-of-mass energies $\sqrt{s}=3.46$
and 5.29 TeV. In addition, with the increase in the center-of-mass energy of the collider of 3.46 to 5.29 TeV, the cross-section increases
by up to 2 orders of magnitude. The results shown in Tables VIII and IX are consistent with those of Figs. 7-10.

\begin{table} [h]
\caption{ The total cross-sections of the $e^-p \to e^-\gamma^*\gamma^*p \to e^-W^+W^-p$ signal for $\sqrt{s}=3.46, 5.29$ TeV
at the FCC-he depending on thirteen anomalous couplings obtained by dimension-8 operators. The cross-section for each
coupling are calculated while fixing the other couplings to zero.
The pure-leptonic decay channel of the $W^+W^-$ in the final state are considered.}
\begin{tabular}{|c|c|c|}
\hline\hline
Couplings (TeV$^{-4}$)  &  $\sigma_{ep} (\mbox{pb})@ \sqrt{s}=3.46$ TeV &  $\sigma_{ep}(\mbox{pb})@ \sqrt{s}=5.29$ TeV    \\
\hline \hline
$f_{M0}/\Lambda^{4}$  & 7.54 $\times 10^{-1}$ & 5.38 $\times 10^{1}$ \\
\hline
$f_{M1}/\Lambda^{4}$  & 6.40 $\times 10^{-2}$ & 4.13  \\
\hline
$f_{M2}/\Lambda^{4}$  & 3.24 $\times 10^{1}$ & 2.31 $\times 10^{3}$ \\
\hline
$f_{M3}/\Lambda^{4}$  & 2.56  & 1.77 $\times 10^{2}$ \\
\hline
$f_{M4}/\Lambda^{4}$  & 2.47  & 1.76 $\times 10^{2}$ \\
\hline
$f_{M5}/\Lambda^{4}$  & 1.93 $\times 10^{-1}$  & 1.35 $\times 10^{1}$ \\
\hline
$f_{M7}/\Lambda^{4}$  & 1.66 $\times 10^{-2}$  & 1.03 \\
\hline
$f_{T0}/\Lambda^{4}$  &  2.50 $\times 10^{1}$ & 1.71 $\times 10^{3}$  \\
\hline
$f_{T1}/\Lambda^{4}$  &  1.62 & 1.02 $\times 10^{2}$  \\
\hline
$f_{T2}/\Lambda^{4}$  &  1.75 & 1.17 $\times 10^{2}$ \\
\hline
$f_{T5}/\Lambda^{4}$  & 2.68 $\times 10^{2}$ & 1.83 $\times 10^{4}$   \\
\hline
$f_{T6}/\Lambda^{4}$  & 1.73 $\times 10^{1}$  & 1.10 $\times 10^{3}$  \\
\hline
$f_{T7}/\Lambda^{4}$  & 1.85 $\times 10^{1}$  & 1.26 $\times 10^{3}$  \\
\hline
\end{tabular}
\end{table}

\begin{table} [h]
\caption{ The total cross-sections of the $e^-p \to e^-\gamma^*\gamma^*p \to e^-W^+W^-p$ signal for $\sqrt{s}=3.46, 5.29$ TeV
at the FCC-he depending on thirteen anomalous couplings obtained by dimension-8 operators. The cross-section for each
coupling are calculated while fixing the other couplings to zero.
The semi-leptonic decay channel of the $W^+W^-$ in the final state are considered.}
\begin{tabular}{|c|c|c|}
\hline\hline
Couplings (TeV$^{-4}$)  & $\sigma_{ep} (\mbox{pb})@ \sqrt{s}=3.46$ TeV &  $\sigma_{ep}(\mbox{pb})@ \sqrt{s}=5.29$ TeV    \\
\hline \hline
$f_{M0}/\Lambda^{4}$  & 1.10 $\times 10^{1}$ & 7.45 $\times 10^{1}$ \\
\hline
$f_{M1}/\Lambda^{4}$  & 7.73 $\times 10^{-1}$ & 5.00 \\
\hline
$f_{M2}/\Lambda^{4}$  & 4.75 $\times 10^{2}$  & 3.18 $\times 10^{3}$ \\
\hline
$f_{M3}/\Lambda^{4}$  & 3.23 $\times 10^{1}$  & 2.12 $\times 10^{2}$ \\
\hline
$f_{M4}/\Lambda^{4}$  & 3.61 $\times 10^{1}$  & 2.43 $\times 10^{2}$ \\
\hline
$f_{M5}/\Lambda^{4}$  & 2.45  & 1.61 $\times 10^{1}$ \\
\hline
$f_{M7}/\Lambda^{4}$  & 1.95 $\times 10^{-1}$  & 1.25 \\
\hline
$f_{T0}/\Lambda^{4}$  &  4.07 $\times 10^{2}$ & 3.62 $\times 10^{3}$  \\
\hline
$f_{T1}/\Lambda^{4}$  &  5.77 $\times 10^{1}$ & 5.40 $\times 10^{2}$  \\
\hline
$f_{T2}/\Lambda^{4}$  &  3.26 $\times 10^{1}$ & 2.91 $\times 10^{2}$  \\
\hline
$f_{T5}/\Lambda^{4}$  & 4.39 $\times 10^{3}$  & 3.90 $\times 10^{4}$   \\
\hline
$f_{T6}/\Lambda^{4}$  & 6.17 $\times 10^{2}$  & 5.82 $\times 10^{3}$  \\
\hline
$f_{T7}/\Lambda^{4}$  & 3.49 $\times 10^{2}$  & 3.15 $\times 10^{3}$  \\
\hline
\end{tabular}
\end{table}

In Tables X-XIII, we list the estimated bounds at $95\%$ C.L. on the full set of dimension-8 coefficients consider in this paper.
In addition, our limits for $f_{M,i}/\Lambda^{4}$ and $f_{T,i}/\Lambda^{4}$ parameters are model-independent,
and the tables also show that the most stringent at $95\%$ C. L. limits are obtained for $f_{T,5}/\Lambda^{4}$, $f_{M,2}/\Lambda^{4}$
and $f_{T,0}/\Lambda^{4}$, respectively. In general, our results reported in Tables X-XIII are approximately one order of magnitude
stringent than the CMS Collaboration limits (see Table I), which indicate the impact that the $e^-p \to e^-\gamma^*\gamma^*p \to e^-W^+W^-p$
process, as well as the cleaner environments that the FCC-he can have on our results. Furthermore, our results compare favorably with other
results reported in the literature.

\begin{table} [h]
\caption{Limits at $95\%$ C.L. on the anomalous $WW\gamma\gamma$ quartic couplings of the $e^-p \to e^-\gamma^*\gamma^*p \to e^-W^+W^-p$
signal for $\sqrt{s}=3.46$ TeV at the FCC-he. The coupling are calculated while fixing the other couplings to zero.
The pure-leptonic decay channel of the $W^+W^-$ in the final state are considered.}
\begin{tabular}{|c|c|c|c|c|c|}
\hline\hline
Couplings (TeV$^{-4}$) & 100 fb$^{-1}$ & 300 fb$^{-1}$ & 500 fb$^{-1}$ & 1000 fb$^{-1}$ \\
\hline \hline
$f_{M0}/\Lambda^{4}$  & [-0.99; 1.10] $\times 10^{2}$ & [-0.12; 0.12] $\times 10^{2}$ & [-0.10; 0.11] $\times 10^{2}$ & [-0.86;0.91] $\times 10^{1}$ \\
\hline
$f_{M1}/\Lambda^{4}$  & [-0.46; 0.30] $\times 10^{3}$ & [-0.37; 0.22]$ \times 10^{3}$  & [-0.34; 0.19] $ \times 10^{3}$ & [-0.30; 0.15] $ \times 10^{3}$   \\
\hline
$f_{M2}/\Lambda^{4}$  & [-0.15; 0.16] $\times 10^{2}$ & [-0.12; 0.13]$ \times 10^{2}$  & [-0.10; 0.11] $ \times 10^{2}$ & [-0.87; 0.93]$\times 10^{1}$   \\
\hline
$f_{M3}/\Lambda^{4}$  & [-0.66; 0.50] $\times 10^{2}$ & [-0.52; 0.36]$ \times 10^{2}$  & [-0.47; 0.31] $ \times 10^{2}$ & [-0.41; 0.25]$\times 10^{2}$   \\
\hline
$f_{M4}/\Lambda^{4}$  & [-0.54; 0.62] $\times 10^{2}$ & [-0.40; 0.49]$ \times 10^{2}$  & [-0.34; 0.43] $ \times 10^{2}$ & [-0.28; 0.37]$\times 10^{2}$   \\
\hline
$f_{M5}/\Lambda^{4}$  & [-0.25; 0.17] $\times 10^{3}$ & [-0.20; 0.12]$ \times 10^{3}$  & [-0.18; 0.10] $ \times 10^{3}$ & [-0.16; 0.08]$\times 10^{3}$   \\
\hline
$f_{M7}/\Lambda^{4}$  & [-0.60; 0.93]$\times 10^{3}$    & [-0.43; 0.75]$\times 10^{3}$   & [-0.36; 0.69]$\times 10^{3}$   & [-0.29; 0.61]$\times 10^{3}$    \\
\hline
$f_{T0}/\Lambda^{4}$  &  [-0.22; 0.15] $\times 10^{2}$ & [-0.17; 0.11] $\times 10^{2}$ & [-0.16; 0.09] $\times 10^{2}$ & [-0.14; 0.08] $\times 10^{2}$ \\
\hline
$f_{T1}/\Lambda^{4}$  & [-0.99; 0.48] $\times 10^{2}$  & [-0.84; 0.33] $\times 10^{2}$ & [-0.78; 0.27] $\times 10^{2}$ & [-0.72; 0.21] $\times 10^{2}$  \\
\hline
$f_{T2}/\Lambda^{4}$  &  [-0.91; 0.56] $\times 10^{2}$ & [-0.75; 0.40] $\times 10^{2}$ & [-0.69; 0.34] $\times 10^{2}$ & [-0.61; 0.26] $\times 10^{2}$ \\
\hline
$f_{T5}/\Lambda^{4}$  & [-0.56; 0.55] $\times 10^{1}$  & [-0.43; 0.42] $\times 10^{1}$ & [-0.38; 0.37] $\times 10^{1}$ & [-0.32; 0.31] $\times 10^{1}$  \\
\hline
$f_{T6}/\Lambda^{4}$  & [-0.25; 0.19] $\times 10^{2}$  & [-0.20; 0.14] $\times 10^{2}$ & [-0.18; 0.12] $\times 10^{2}$ & [-0.16; 0.09] $\times 10^{2}$ \\
\hline
$f_{T7}/\Lambda^{4}$  & [-0.26; 0.17] $\times 10^{2}$  & [-0.21; 0.12] $\times 10^{2}$ & [-0.19; 0.10] $\times 10^{2}$ & [-0.17; 0.08] $\times 10^{2}$ \\
\hline
\end{tabular}
\end{table}

\begin{table} [h]
\caption{Limits at $95\%$ C.L. on the anomalous $WW\gamma\gamma$ quartic couplings of the $e^-p \to e^-\gamma^*\gamma^*p \to e^-W^+W^-p$
signal for $\sqrt{s}=5.29$ TeV at the FCC-he. The coupling are calculated while fixing the other couplings to zero.
The pure-leptonic decay channel of the $W^+W^-$ in the final state are considered.}
\begin{tabular}{|c|c|c|c|c|c|}
\hline\hline
Couplings (TeV$^{-4}$) & 100 fb$^{-1}$ & 300 fb$^{-1}$ & 500 fb$^{-1}$ & 1000 fb$^{-1}$ \\
\hline \hline
$f_{M0}/\Lambda^{4}$  & [-0.15; 0.16] $\times 10^{2}$ & [-0.11; 0.12] $\times 10^{2}$ & [-0.10; 0.11] $\times 10^{2}$ & [-0.86; 0.91] $\times 10^{1}$ \\
\hline
$f_{M1}/\Lambda^{4}$  & [-0.57; 0.56] $\times 10^{2}$ & [-0.44; 0.43]$ \times 10^{2}$  & [-0.38; 0.37] $ \times 10^{2}$ & [-0.32; 0.31] $ \times 10^{2}$   \\
\hline
$f_{M2}/\Lambda^{4}$  & [-0.23; 0.25] $\times 10^{1}$ & [-0.17; 0.19]$ \times 10^{1}$  & [-0.15; 0.17] $ \times 10^{1}$ & [-0.12; 0.15]$\times 10^{1}$   \\
\hline
$f_{M3}/\Lambda^{4}$  & [-0.88; 0.85] $\times 10^{1}$ & [-0.67; 0.64]$ \times 10^{1}$  & [-0.59; 0.56] $ \times 10^{1}$ & [-0.50; 0.47]$\times 10^{1}$   \\
\hline
$f_{M4}/\Lambda^{4}$  & [-0.85; 0.88] $\times 10^{1}$ & [-0.64; 0.67]$ \times 10^{1}$  & [-0.57; 0.59] $ \times 10^{1}$ & [-0.47; 0.50]$\times 10^{1}$   \\
\hline
$f_{M5}/\Lambda^{4}$  & [-0.32; 0.31] $\times 10^{2}$ & [-0.25; 0.23]$ \times 10^{2}$  & [-0.22; 0.20] $ \times 10^{2}$ & [-0.18; 0.17]$\times 10^{2}$   \\
\hline
$f_{M7}/\Lambda^{4}$  & [-0.11; 0.12]$\times 10^{3}$    & [-0.84; 0.89]$\times 10^{2}$   & [-0.74; 0.78]$\times 10^{2}$   & [-0.62; 0.66]$\times 10^{2}$    \\
\hline
$f_{T0}/\Lambda^{4}$  &  [-0.30; 0.26] $\times 10^{1}$ & [-0.24; 0.19] $\times 10^{1}$ & [-0.21; 0.16] $\times 10^{1}$ & [-0.18; 0.14] $\times 10^{1}$ \\
\hline
$f_{T1}/\Lambda^{4}$  & [-0.13; 0.10] $\times 10^{2}$  & [-0.10; 0.07] $\times 10^{2}$ & [-0.91; 0.63] $\times 10^{1}$ & [-0.80; 0.51] $\times 10^{1}$  \\
\hline
$f_{T2}/\Lambda^{4}$  &  [-0.11; 0.10] $\times 10^{2}$ & [-0.87; 0.75] $\times 10^{1}$ & [-0.77; 0.66] $\times 10^{1}$ & [-0.66; 0.54] $\times 10^{1}$ \\
\hline
$f_{T5}/\Lambda^{4}$  & [-1.01; 0.72]  & [-0.81; 0.52] & [-0.73; 0.44] & [-0.64; 0.36]  \\
\hline
$f_{T6}/\Lambda^{4}$  & [-0.38; 0.32] $\times 10^{1}$  & [-0.30; 0.23] $\times 10^{1}$ & [-0.26; 0.20] $\times 10^{1}$ & [-0.23; 0.17] $\times 10^{1}$ \\
\hline
$f_{T7}/\Lambda^{4}$  & [-0.33; 0.32] $\times 10^{1}$  & [-0.25; 0.24] $\times 10^{1}$ & [-0.22; 0.21] $\times 10^{1}$ & [-0.19; 0.18] $\times 10^{1}$ \\
\hline
\end{tabular}
\end{table}

\begin{table} [h]
\caption{Limits at $95\%$ C.L. on the anomalous $WW\gamma\gamma$ quartic couplings of the $e^-p \to e^-\gamma^*\gamma^*p \to e^-W^+W^-p$
signal for $\sqrt{s}=3.46$ TeV at the FCC-he. The coupling are calculated while fixing the other couplings to zero.
The semi-leptonic decay channel of the $W^+W^-$ in the final state are considered.}
\begin{tabular}{|c|c|c|c|c|c|}
\hline\hline
Couplings (TeV$^{-4}$) & 100 fb$^{-1}$ & 300 fb$^{-1}$ & 500 fb$^{-1}$ & 1000 fb$^{-1}$ \\
\hline \hline
$f_{M0}/\Lambda^{4}$  & [-0.37; 0.41] $\times 10^{2}$ & [-0.27; 0.31] $\times 10^{2}$ & [-0.24; 0.28] $\times 10^{2}$ & [-0.20; 0.24] $\times 10^{1}$ \\
\hline
$f_{M1}/\Lambda^{4}$  & [-0.17; 0.13] $\times 10^{3}$ & [-0.14; 0.09]$ \times 10^{3}$  & [-0.13; 0.08] $ \times 10^{3}$ & [-0.11; 0.06] $ \times 10^{3}$   \\
\hline
$f_{M2}/\Lambda^{4}$  & [-0.57; 0.60] $\times 10^{1}$ & [-0.43; 0.46]$ \times 10^{1}$  & [-0.37; 0.41] $ \times 10^{1}$ & [-0.31; 0.35]$\times 10^{1}$   \\
\hline
$f_{M3}/\Lambda^{4}$  & [-0.25; 0.20] $\times 10^{2}$ & [-0.20; 0.15]$ \times 10^{2}$  & [-0.18; 0.13] $ \times 10^{2}$ & [-0.16; 0.10]$\times 10^{2}$   \\
\hline
$f_{M4}/\Lambda^{4}$  & [-0.20; 0.22] $\times 10^{2}$ & [-0.15; 0.17]$ \times 10^{2}$  & [-0.13; 0.15] $ \times 10^{2}$ & [-0.11; 0.13]$\times 10^{2}$   \\
\hline
$f_{M5}/\Lambda^{4}$  & [-0.92; 0.72] $\times 10^{2}$ & [-0.72; 0.53]$ \times 10^{2}$  & [-0.65; 0.45] $ \times 10^{2}$ & [-0.57; 0.37]$\times 10^{2}$   \\
\hline
$f_{M7}/\Lambda^{4}$  & [-0.25; 0.35]$\times 10^{3}$    & [-0.18; 0.28]$\times 10^{3}$   & [-0.15; 0.25]$\times 10^{3}$   & [-0.12; 0.22]$\times 10^{3}$    \\
\hline
$f_{T0}/\Lambda^{4}$  &  [-0.93; 0.43] $\times 10^{1}$ & [-0.80; 0.29] $\times 10^{1}$ & [-0.75; 0.24] $\times 10^{1}$ & [-0.69; 0.18] $\times 10^{1}$ \\
\hline
$f_{T1}/\Lambda^{4}$  & [-0.16;0.18] $\times 10^{2}$  & [-0.17; 0.14] $\times 10^{2}$ & [-0.10; 0.12] $\times 10^{2}$ & [-0.08; 0.11] $\times 10^{2}$  \\
\hline
$f_{T2}/\Lambda^{4}$  &  [-0.35; 0.14] $\times 10^{2}$ & [-0.30; 0.10] $\times 10^{2}$ & [-0.29; 0.08] $\times 10^{2}$ & [-0.27; 0.06] $\times 10^{2}$ \\
\hline
$f_{T5}/\Lambda^{4}$  & [-0.20; 0.18] $\times 10^{1}$  & [-0.16; 0.14] $\times 10^{1}$ & [-0.14; 0.12] $\times 10^{1}$ & [-0.12; 0.10] $\times 10^{1}$  \\
\hline
$f_{T6}/\Lambda^{4}$  & [-0.65; 0.41] $\times 10^{1}$  & [-0.53; 0.29] $\times 10^{1}$ & [-0.49; 0.24] $\times 10^{1}$ & [-0.43; 0.19] $\times 10^{1}$ \\
\hline
$f_{T7}/\Lambda^{4}$  & [-0.71; 0.66] $\times 10^{1}$  & [-0.55; 0.49] $\times 10^{1}$ & [-0.49; 0.43] $\times 10^{1}$ & [-0.41; 0.36] $\times 10^{1}$ \\
\hline
\end{tabular}
\end{table}

\begin{table} [h]
\caption{Limits at $95\%$ C.L. on the anomalous $WW\gamma\gamma$ quartic couplings of the $e^-p \to e^-\gamma^*\gamma^*p \to e^-W^+W^-p$
signal for $\sqrt{s}=5.29$ TeV at the FCC-he. The coupling are calculated while fixing the other couplings to zero.
The semi-leptonic decay channel of the $W^+W^-$ in the final state are considered.}
\begin{tabular}{|c|c|c|c|c|c|}
\hline\hline
Couplings (TeV$^{-4}$) & 100 fb$^{-1}$ & 300 fb$^{-1}$ & 500 fb$^{-1}$ & 1000 fb$^{-1}$ \\
\hline \hline
$f_{M0}/\Lambda^{4}$  & [-0.17; 0.19] $\times 10^{2}$ & [-0.13; 0.15] $\times 10^{2}$ & [-0.11; 0.13] $\times 10^{2}$ & [-0.09; 0.11] $\times 10^{1}$ \\
\hline
$f_{M1}/\Lambda^{4}$  & [-0.81; 0.60] $\times 10^{2}$ & [-0.64; 0.44]$ \times 10^{2}$  & [-0.58; 0.37] $ \times 10^{2}$ & [-0.51; 0.30] $ \times 10^{2}$   \\
\hline
$f_{M2}/\Lambda^{4}$  & [-0.26; 0.29] $\times 10^{1}$ & [-0.19; 0.23]$ \times 10^{1}$  & [-0.17; 0.20] $ \times 10^{1}$ & [-0.14; 0.17]$\times 10^{1}$   \\
\hline
$f_{M3}/\Lambda^{4}$  & [-0.11; 0.10] $\times 10^{2}$ & [-0.82; 0.79]$ \times 10^{1}$  & [-0.73; 0.70] $ \times 10^{1}$ & [-0.61; 0.58]$\times 10^{1}$   \\
\hline
$f_{M4}/\Lambda^{4}$  & [-0.98; 1.00] $\times 10^{1}$ & [-0.75; 0.76]$ \times 10^{1}$  & [-0.66; 0.67] $ \times 10^{1}$ & [-0.55; 0.56]$\times 10^{1}$   \\
\hline
$f_{M5}/\Lambda^{4}$  & [-0.42; 0.35] $\times 10^{2}$ & [-0.33; 0.26]$ \times 10^{2}$  & [-0.29; 0.23] $ \times 10^{2}$ & [-0.25; 0.19]$\times 10^{2}$   \\
\hline
$f_{M7}/\Lambda^{4}$  & [-0.12; 0.17]$\times 10^{3}$    & [-0.85; 1.32]$\times 10^{2}$   & [-0.72; 1.20]$\times 10^{2}$   & [-0.58; 1.06]$\times 10^{2}$    \\
\hline
$f_{T0}/\Lambda^{4}$  &  [-0.33; 0.20] $\times 10^{1}$ & [-0.27; 0.14] $\times 10^{1}$ & [-0.25; 0.12] $\times 10^{1}$ & [-0.23; 0.09] $\times 10^{1}$ \\
\hline
$f_{T1}/\Lambda^{4}$  & [-0.73; 0.61] $\times 10^{1}$  & [-0.57; 0.45] $\times 10^{1}$ & [-0.51; 0.39] $\times 10^{1}$ & [-0.44; 0.32] $\times 10^{1}$  \\
\hline
$f_{T2}/\Lambda^{4}$  &  [-0.10; 0.08] $\times 10^{2}$ & [-0.79; 0.60] $\times 10^{1}$ & [-0.70; 0.52] $\times 10^{1}$ & [-0.61; 0.42] $\times 10^{1}$ \\
\hline
$f_{T5}/\Lambda^{4}$  & [-0.83; 0.73]  & [-0.65; 0.55] & [-0.58; 0.48] & [-0.49; 0.39]  \\
\hline
$f_{T6}/\Lambda^{4}$  & [-0.26; 0.16] $\times 10^{1}$  & [-0.21; 0.12] $\times 10^{1}$ & [-0.19; 0.10] $\times 10^{1}$ & [-0.17; 0.08] $\times 10^{1}$ \\
\hline
$f_{T7}/\Lambda^{4}$  & [-0.31; 0.25] $\times 10^{1}$  & [-0.24; 0.18] $\times 10^{1}$ & [-0.22; 0.16] $\times 10^{1}$ & [-0.19; 0.13] $\times 10^{1}$ \\
\hline
\end{tabular}
\end{table}

\section{Conclusions}

Vector boson scattering processes are widely recognized as the best laboratory to study the operators which modify only the quartic $VVVV$ couplings.
The studies of the anomalous $WW\gamma\gamma$ coupling through the sub-process $\gamma^{*}\gamma^{*}  \to W^+W^-$ at $pp$ and $e^+e^-$ colliders were investigated by Refs.\cite{Chapon,Koksal2}. Secondly, the sub-process $\gamma^{*}\gamma^{*}  \to W^+W^-$ includes only interactions between the gauge bosons, causing more apparent possible deviations from the expected value of SM. Also, an important advantage of the process $e^-p \to e^-\gamma^*\gamma^*p \to e^-W^+W^-p$ at the LHeC and the FCC-he is that it isolates $WW \gamma \gamma$ coupling from the other quartic couplings as seen from the Feynman diagrams. Thus, the anomalous $WW \gamma \gamma$ coupling can be studied alone by means of the sub-process $\gamma^*\gamma^* \to W^+W^-$ at $ep$ colliders.

For these reasons, the parameterization of new physics effects in $e^-p \to e^-\gamma^*\gamma^*p \to e^-W^+W^-p$ scattering via $\gamma^*\gamma^* \rightarrow W^+W^-$ is a useful tool for analyzing BSM at the LHeC and the FCC-he.

In conclusion, our results reported in this paper through Figs. 3-10 and Tables II-XIII on the total cross-section for the
$e^-p \to e^-\gamma^*\gamma^*p \to e^-W^+W^-p$ process, as well as of the aQGC $f_{M,i}/\Lambda^{4}$ and $f_{T,i}/\Lambda^{4}$,
show that the $e^-p \to e^-\gamma^*\gamma^*p \to e^-W^+W^-p$ process is a very good option to measured the total cross-section
and to probing the aQGC at the LHeC and the FCC-he with good sensitivity.

\vspace{10mm}

\begin{center}
{\bf Acknowledgments}
\end{center}

A. G. R. and M. A. H. R. thank SNI and PROFEXCE (M\'exico).

\vspace{2cm}


\newpage

\end{document}